\documentclass[5p,a4paper, sorted]{elsarticle}

\usepackage{subcaption}
\usepackage{lineno, amsmath, graphicx}
\usepackage[draft]{hyperref}

\graphicspath{{Pictures/}}

\journal{Computer Physics Communications}

\usepackage{soul}

\usepackage{amsmath}
\usepackage{amssymb}
\usepackage{color}
\usepackage{nicefrac}

\renewcommand{\kappa}{\varkappa}

\usepackage{algorithm}
\usepackage{algpseudocode}

\makeatletter
\newcommand{\vvec}[1]{\mathbf{#1}}
\makeatother

\newcommand{\VD}{VD$^3$ }

\begin{document}

\begin{frontmatter}

\title{Parallel SPH modeling using dynamic domain decomposition and load balancing displacement of Voronoi subdomains}

\address[dukhov]{Dukhov Research Institute of Automatics (VNIIA), Moscow, Russia}
\address[mipt]{Moscow Institute of Physics and Technology, Dolgoprudny, Moscow Region, Russia}
\address[jiht]{Joint Institute for High Temperatures of the Russian Academy of Sciences, Moscow, Russia}
\address[landau]{Landau Institute for Theoretical Physics, Chernogolovka, Russia}
\author[dukhov,landau]{M. S. Egorova}
\ead{egorova.maria.serg@gmail.com}
\author[dukhov,mipt,jiht,landau]{S. A. Dyachkov}
\author[dukhov,jiht]{A. N. Parshikov}
\author[dukhov,landau]{V. V. Zhakhovsky}
\ead{vasily@mpd3.com}

\begin{abstract}
A highly adaptive load balancing algorithm for parallel simulations using particle methods, such as molecular dynamics and smoothed particle hydrodynamics (SPH), is developed. Our algorithm is based on the dynamic spatial decomposition of simulated material samples between Voronoi subdomains, where each subdomain with all its particles is handled by a~single computational process which is typically run on a~single CPU core of a~multiprocessor computing cluster.

The algorithm displaces the positions of neighbor Voronoi subdomains in accordance with the local load imbalance between the corresponding processes. It results in particle transfers from heavy-loaded processes to less-loaded ones. Iteration of the algorithm puts into alignment the processor loads. Convergence to a~well-balanced decomposition from imbalanced one is improved by the usage of multi-body terms in the balancing displacements.

The high adaptability of the balancing algorithm to simulation conditions is illustrated by SPH modeling of the dynamic behavior of materials under extreme conditions, which are characterized by large pressure and velocity gradients, as a~result of which the spatial distribution of particles varies greatly in time. The~higher parallel efficiency of our algorithm in such conditions is demonstrated by comparison with the corresponding static decomposition of the computational domain. Our algorithm shows almost perfect strong scalability in tests using from tens to several thousand processes.
\end{abstract}


\end{frontmatter}


\section{Introduction}

The unsteady motion of materials in extreme conditions is characterized by significant pressure and velocity gradients, as well as the occurrence of free surfaces and contact density discontinuities. Efficient parallel modeling of such motion encounters a~problem of load balancing between computational resources in use. Algorithms using a~static spatial decomposition of the computational domain cannot provide fair parallel efficiency, since the distribution of the~workload, determined by the computational time required for calculating the motion, is not tied to the unsteady material flow. The~existing load balancing algorithms discussed below are also inadequate to adapt to rapid changes in spatial distribution of material.

This work aims to develop a~highly adaptable load balancing algorithm which can be utilized in a~ parallel program designed for multiprocessor computing clusters. The~efficient program must be able to load available processors with useful work evenly and maximally, thereby ensuring the high parallel efficiency of hydrodynamics simulation of high-rate processes and high-speed motion produced in materials under extreme conditions. Such conditions are realized in impact on heterogeneous obstacles, shock wave propagation in mesostructures of porous material, and other challenging problems which call for efficient parallel codes.

Efficient parallel modeling of detonation of explosives with pores and/or inert additives, splitting and merging of droplets in high-speed flows, impact fragmentation of brittle materials and high-speed collision of bodies is also demanded. The~computational challenge of simulating such processes can be naturally addressed by employing meshless particle methods, such as molecular dynamics (MD) and smoothed particle hydrodynamics (SPH) methods, implemented in efficient parallel programs.

To model the processes listed above the meshless SPH method~\cite{monaghan1977smoothed} is preferable to the grid methods that encounter great difficulties in the exact resolution of shock fronts, contact surfaces and free boundaries (for the Euler formalism) and the construction of movable adaptive grids under the Lagrangian representation of continuum. In this paper a~further description is based on our experience in developing a~massively parallel program for the contact SPH method (CSPH) utilizing the Riemann solver~\cite{parshikov2002smoothed} for interparticle interaction. However, an algorithm of dynamic domain decomposition described below is quite universal and can be applied not only for SPH, but it is suitable for any other particle method with a~finite radius of interaction between particles; for example, for the direct Monte-Carlo, particle in cell (PIC) and MD methods.

The optimal distribution of computational loads between threads, computational processes or processor cores on a~computing cluster is the cornerstone of efficient parallel program architecture. For the hydrodynamic modeling, the problem of load assignment reduces to a~partitioning of samples, i.e., to the decomposition of simulated material in samples between computational processes. An even decomposition is necessary, but not a~sufficient condition for optimality, since the data exchange between processes is determined by numbers of interacting particles that are found in different processes. These particles are distributed in the boundary layers between the processes if the spatial decomposition on subdomains among the processes is used. The~exchange of information on boundary particles is required to support interaction between neighboring subdomains that belong to different processes.

As it is shown in~\cite{plimpton1995fast}, the spatial decomposition algorithms for MD method are optimal among other concepts of decomposition (in comparison with decomposition by particle numbers~\cite{plimpton1995fast} or decomposition of a~matrix of forces~\cite{murty1999efficient, shu2003optimization}), as they allow not only to equalize the workload between processes, but also to minimize the data exchange between them. An extensive data exchange, which is required in the decomposition by particle numbers for taking into account interaction of particles randomly distributed in the samples, does not allow achieving the high computational efficiency on large clusters with limited network bandwidth.

The most straightforward method of spatial decomposition is a~partitioning of a~computational domain into equal rectangular subdomains~\cite{zhang2009fast, cherfils2012josephine}. In the context of spatial nonuniformity of particle concentration, the partitioning can be performed recursively, using the orthogonal recursive bisection (ORB) method. General issues of applying this approach to particle methods with a~finite interaction radius were considered in~\cite{fleissner2008parallel}, the specific case of its application to SPH was presented in~\cite{Sbalzarini2006ppm,oger2016distributed}. The~idea of the technique is simple: a~rectangular area is recursively divided into two parts along the long side so that an equal number of particles are found in each. The~boundaries between rectangular subdomains are planes that are parallel to coordinate axes. Balancing is carried out via mobility of the planes separating those subdomains. But with re-balancing, the connectivity between processes can change significantly. Since subdomains have different numbers of neighbors, a~reorganization of the decomposition requires the extensive particle exchanges. This decomposition method also does not take into account particle mixing during their motion. All this, taken together, increases the calculation time due to extensive particle data exchanges in the boundary layers between the processes.

An approach that uses decomposition based on rectangles generated not from a~large scale to a~smaller one like it is done by ORB, but from a~small scale to a~larger scale is also possible. The~computational domain can be decomposed by a~Cartesian grid into small cells, and the decomposition is performed by distributing the groups of microcells among the processes~\cite{sato2006implementing, ferrari2009new}. Microcells of one process must form a~simply connected region. Microcells can migrate between processes, but the connectivity of the macro-regions remains constant. Such a~limitation is a~significant drawback for our problems, where the particles can significantly change the spatial arrangement because the undesirable additional particle data exchanges between the processes are required due to the migration of microcells mentioned above.

The third, more complex type of decomposition is the generation of an auxiliary grid in the computational domain, the cells of which, together with the particles in cells, are distributed among the computational processes~\cite{deng2000adaptive, begau2015adaptive}. Load balancing is possible throughout the mobility of grid nodes, the connectivity of which is usually assumed to be fixed. During the calculation, the grid nodes are shifted towards the ``load center'' of the cells that have this node in the vertex list. The~load center is approximated, for example, by the center of mass of the cells containing particles. To save the grid structure, the global decomposition adjustment algorithm is used. The~main disadvantage of this decomposition method in application to our problems is that the grid cells can become significantly deformed, which increases the exchange of particles due to the lengthening of the cell boundaries and growth of the contact surface of neighboring cells.

Many works were presented in an attempt to find the most efficient technology of particle redistribution. The~procedure for updating the decomposition should be consistent with the load balancing data. It is stated~\cite{cybenko1989dynamic, willebeek1993strategies, hu1998optimal} that the ideal decomposition algorithm must be of diffusion type: the computational load should be redistributed in a~manner similar to the heat conduction process. The~workload is measured either proportionally to a~number of particles in the computational subdomains~\cite{cherfils2012josephine, oger2016distributed, ferrari2009new}, or, more naturally, by the time spent on computations~\cite{zhang2009fast, fleissner2008parallel, zhakhovskii2005new, fattebert2012dynamic}.

The algorithm we propose is designed for massively parallel systems with distributed memory. It uses the idea of dynamic Voronoi decomposition~\cite{zhakhovskii2005new, fattebert2012dynamic, koradi2000point}, which is combined with some load auto-balancing algorithm. In accordance with~\cite{cybenko1989dynamic, willebeek1993strategies, hu1998optimal}, the load can be measured as a~ratio of the useful calculation time to the total wall clock time required to accomplish a~simulation step. We use this definition of load and adopt the method of load balancing for the Voronoi dynamic domain decomposition \VD initially proposed for MD method in~\cite{zhakhovskii2005new}. Another definitions of load can be also utilized, in particular the number of inter-particle interactions calculated in each Voronoi domain or the number of sent/received data, but aligning such loads may not correlate with the real processor loads in highly inhomogeneous flows.

Voronoi decomposition is uniquely defined by a~set of generator points of the Voronoi diagram, where each point is associated with its cell and handled by a~single process. The~particles nearest to the Voronoi cell generator point are stored in memory and updated by a~corresponding process. The~mobility of the Voronoi diagram is realized through a~mobility of the point generators because their positions are bound to movement of their own particles. The~process load for each Voronoi cell can vary in time, which is caused by the following factors. First, the number of particles in the cell can vary due to algorithmic reasons (splitting and merging of particles, correction of the boundaries of the diagram due to displacements of the centers of the diagram), as well as to physical reasons including changes of density distribution and/or disruption of the continuity in media surrounding generator point. Secondly, the physical processes inside all particle are local, and therefore can be calculated by their own algorithms for different times. The~above factors lead to the fact that sets of SPH particles in each Voronoi subdomain are processed for different times, which means an uneven load of processes. Therefore, the positions of the point generators are further adjusted by the balancing displacements resulting in particle transfer from heavily load cells to lesser load cells, which balances the process loads.

The balancing displacement can be expressed in various ways, but the known formulations utilize only the two-body terms describing the pair contributions of neighbor Voronoi subdomains~\cite{zhakhovskii2005new, fattebert2012dynamic, koradi2000point}.
Such displacement based merely on two-body contributions does not take into account the presence of other neighbor subdomains. In other words any two-body displacement is a~rough approximation of the best balancing displacement which must depend on all Voronoi subdomains involved in decomposition.

Here we propose a~new expression for balancing displacement using the angle-dependent multi-body terms, which speed up the convergence to a~well-balanced decomposition. It includes the two-body terms and either the tree-body terms for plane decomposition or the four-body terms for volumetric decomposition.

This multi-body auto-balancing algorithm for the \VD combined with the SPH method is implemented in our parallel CSPH code, which is proved to be highly efficient in several tests. The~used algorithm for \VD makes it possible to take into account the redistribution of masses within a~computational domain in a~natural way during modeling. The~algorithm is able to adapt to arbitrary mass flows with minimal exchange of particles between cells without requiring the preservation of connectivity between the diagram generators. The~data exchange between pairs of processes/cells of the diagram is always local. The~load balancing is fully adaptive, and it is not necessary to rebuild it from scratch to maximize the parallel efficiency of simulation. The~cell boundaries through which the particle exchange is performed usually have less area than in the methods with block/rectangle decomposition of the computational domain.

The Voronoi decomposition with this load balancing does not have disadvantages typical for the decomposition methods listed above. Our computational experience shows that the geometry of cells tends to a~honeycomb structure that has the minimum volume of cell boundary layers. It maximizes the loads by reduction of the number of particles to be exchanged. The~Voronoi subdomains also tend to have an equal number of neighboring subdomains while the connectivity of subdomains is not fixed and can change following the material motion. The~changes of the connectivity remain local in this case, that does not require massive communications between all processes because communications between subdomains are carried out only within the circle of the nearest neighborhood. Thus, despite the fact that the Voronoi decomposition is more difficult for programming than the methods mentioned above, it has undeniable advantages in the parallel modeling of complex flows triggered by extreme conditions.

\newcommand{\tu}{t^\text{u}}
\newcommand{\te}{t^\text{e}}
\newcommand{\tw}{t^\text{w}}
\newcommand{\tex}{t^\text{ex}}
\newcommand{\av}[1]{\langle #1 \rangle}

\newcommand{\Dshell}{D^{\text{sh}}}
\newcommand{\Dcell}{D^{\text{c}}}
\newcommand{\Rbuf}{R^{\text{buf}}}
\newcommand{\Rhrz}{R^{\text{hrz}}}
\newcommand{\Rint}{R^{\text{int}}}
\newcommand{\Nupd}{N^{\text{upd}}}

\begin{figure}[t]   
  \center\includegraphics[width=1\columnwidth]{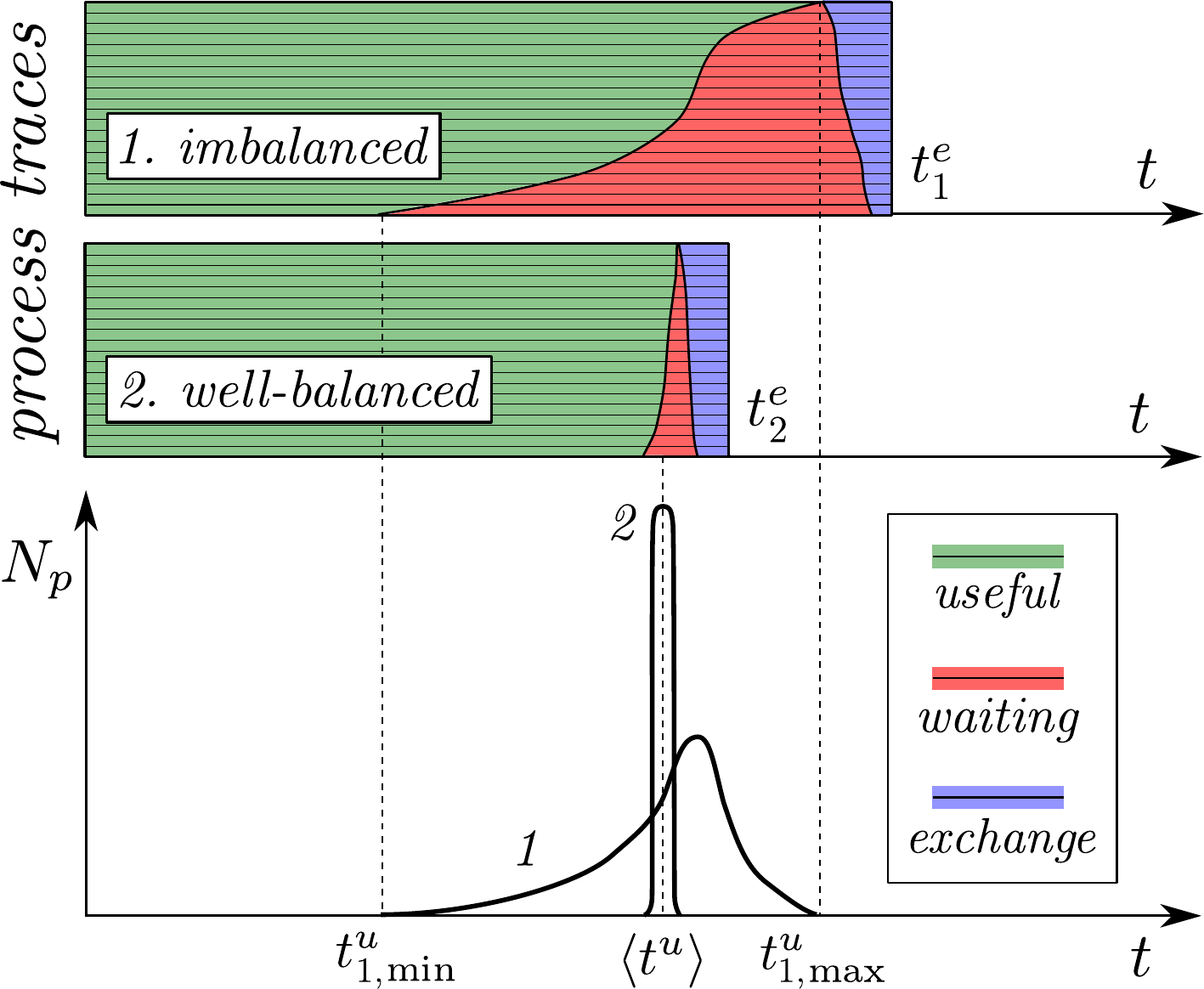} \\
  \caption{The~schematic traces (1 and 2) of the~parallel processes arranged by the~useful times utilized for specific SPH calculations, shown by green lines, within a~simulation step and the~corresponding distributions of the~number of processes as functions of the~useful time. Load balancing~$1\to 2$ leads to a~narrower distribution, which in turn results in shorter waiting times and faster simulation. Here the~sum of useful times and exchange times spent by all processes is assumed fixed for simplicity. In real conditions, the~exchange times are usually reduced with balancing.
}  
\label{fig:stats}
\end{figure}

\section{Simulation time minimization by load balancing}\label{SEC:VoronoiMovement}

Once all particles of simulated samples are distributed among the~processes following the~Voronoi subdomains, simulation of particle motion starts.
Each process that handles particles is hereinafter called as a~worker. Preparation of neighbors list and calculating the~interaction of particles are necessary parts in any particle method implementation, so their execution should be considered as a~useful work utilized locally by each computational process/worker. However, the~parallel algorithms have to carry out additional work on data exchange, which can be also considered as a~necessary work. Having a~well-optimized useful and necessary work controlled essentially by a~total number of particles, the~simulation performance can be improved only by minimizing the times when processes stay in the idle state while waiting for necessary data from other processes. Such minimization of the waiting times leads to reduction of the~elapsed wall clock time of simulation.

Let~$\te$ be an~elapsed time used for one simulation step, which can be easily evaluated in a~program. It is determined by the~barrier synchronization between all worker processes, that is the~workers cannot proceed to the~next step until all workers complete the~current step. This unavoidable synchronization results in the following state $\forall k: \te_k = \te$.

It is also easy to measure the~useful time~$\tu_k$ spent by a~worker~$k$ to perform the inter-particle interactions during one simulation step. The~remaining time~$t^{\text{x}}_k$ is determined through the~difference between the~elapsed time per step~$\te$ and the~useful time:
\begin{equation}\label{eq:times1}
\te = \tu_k + t^\text{x}_k.
\end{equation}

In general, the~communication time~$t^\text{x}_k$ can be split into the idle or waiting time $\tw_k$ and the data exchange time~$\tex_k$, which gives
\begin{equation}\label{eq:times2}
\te = \tu_k + \tex_k + \tw_k,
\end{equation}
Such simplified partitioning the elapsed time is used in Fig.~\ref{fig:stats} where the schematic traces of many parallel processes are presented.

The~parallel modeling efficiency of a~worker~$k$ can be characterized by a~normalized load:
\begin{equation*}
\label{eq:load}
  L_k = \frac{\tu_k}{\te_k}, \quad 0 < L_k < 1,
\end{equation*}
which is called hereafter as a~load of worker~$k$.

In practice the several programs executed simultaneously on same node may compete for CPU resources. Nevertheless, it is usually possible to count the~processor-dependent time which is exclusively spent by CPU on execution of a~particular nonparallel code. Then the~useful work, which usually is the~most resource intensive part, can be evaluated as an~elapsed useful time~$\tu$ or as a~useful CPU-time~$\tu_{\text{CPU}}$. The~fraction~$f^p = \tu_{\text{CPU}} / \tu$ defines the utilized part of computational resource. It can be used to improve the~definition of worker load in a~competitive execution environment as:
\begin{equation*}
  L_k = \frac{\tu_k}{f^p_k \te_k}.
\end{equation*}
This formula can be also applied to take into account the~different computational performances of various CPUs in use, say in the computational grid environments~\cite{zhakhovskii2005new}.

We assume that the~program exclusively runs on a~homogeneous computing cluster without other competing tasks, so~$\forall k$~$f^p_k=1$. The~relation~\eqref{eq:times2} can be rewritten by averaging with the~number of workers:
\begin{equation*}
\te = \av{\tu} + \av{\tex} + \av{\tw}.
\end{equation*}
The~total useful work executed on all involved processes $N_p$ is almost fixed and determined by a~total number of particles involved into particular simulation, so the~average useful time $\av{\tu}=\sum_k\tu_k / N_p=\mathrm{const}$ does not depend on the~distribution of SPH particles among the processes. The~simulation performance can reach its maximum if both $\av{\tw}$ and $\av{\tex}$ are minimized. 

Any spatial decomposition algorithm should aim to minimize the waiting time~$\av{\tw}$ and exchange time~$\av{\tex}$. The~exchange time depends on the~number of particles to be exchanged and the network bandwidth, as well as on the~network latency, but the~latter is assumed to be short enough and not taken into account in this work. Let us assume that the exchange time is a~monotonic function of the number of particles in each process, and the total exchange time is a~constant: $\av{\tex}=\sum_k\tex_k / N_p=\mathrm{const}$.

Thus the~shortest wall-clock time for one step is equal to~$\av{\tu} + \av{\tex}$, which corresponds the~optimal load for all workers ${L}^*_k=\av{\tu}/(\av{\tu} + \av{\tex})$, which can be reached at $\forall k$ $\tw_k\approx 0$. Now it is clear that the net efficiency of parallel modeling increases with reduction of the~time gap between the~heavily loaded workers and a~medium-loaded one. To do this, it is necessary to transfer particles from the heavily loaded processes to the less loaded ones in order to make the~distribution of processes over the~useful time narrower, as illustrated by Fig.~\ref{fig:stats}, which leads to $\av{\tw}\to 0$ and makes the corresponding distributions of both waiting and exchange times narrower as well.

In practice, it is difficult to determine which time is spent for waiting of data transfer, and which for data exchange itself, especially if non-blocking communications are used. The~nonblocking communications makes possible to mask the~waiting time for data transfer:  calculation of all particle pairs pertaining to a~process can be performed while this process is waiting for receiving particles from other processes, see the following  Sections~\ref{subsec:SPHNeighbours}, \ref{subsec:SPHExchange}. Thus, in contrast to exchange and waiting times, the usage of useful time, which is measured easily, for definition of load by the ratio~\eqref{eq:load} is reasonable.

To equalize the loads of all workers involved in modeling we propose the~new multi-body  algorithm for the balancing displacements of Voronoi diagram.

\section{Autobalancing Voronoi dynamic domain decomposition}\label{sec:AllVoronoi}
\subsection{Voronoi domain decomposition}\label{sec:Voronoi}

The~Voronoi diagram according to~\cite{du1999centroidal} is a~decomposition of a~closed subspace~$\bar\Omega\in\mathcal{R}^n$ (a~modeled volume or a~computational domain in an $n$-dimensional space) between~$N_{\hat V}$ subdomains~$\{V_k\}_{k=1}^{N_{\hat V}}$ based on distance to a~specific set of Voronoi generators~$\{G_k\}_{k=1}^{N_{\hat V}}$ points:
\begin{equation}\label{eq:VoronoiDefinition}
   \hat V_k = \{\left. \vvec{x}\in\Omega\right.\,:\,|\vvec{x} - \vvec{g} _k | <| \vvec{x} - \vvec{g} _l |,
   \,l = 1,\, \ldots,\, N_{\hat V}, l \ne k \},
\end{equation}
where~$\vvec{g}_k$ is a~radius-vector of the~point~$G_k$. The~definition~\eqref{eq:VoronoiDefinition} states that a~single Voronoi subdomain is a~convex polytope which contains all points that are closest to its generator with respect to any other generator. The~Voronoi diagram is unique for a~given set~$\{G_k\}_{k=1}^{N_{\hat V}}$.

Suppose there is a~material sample or a~set of samples represented by material particles in a~computational domain, and a~job with~$N_p$ parallel processes can be submitted on a~computing cluster. Let each~$k$-process operates one Voronoi subdomain~$\hat V_k$. A set of generator points~$\{G_k\}_{k=1}^{N_p}$ can be placed someway in this domain, and a~peer-to-peer connection between running processes~$k=1,2,\ldots,N_p$ is established. Then all particles of the~samples can be assigned to different processes according to the~particle positions. That is the~data structures describing particles in the~subdomain~$\hat V_k$ is stored in the~memory of the~process~$k$. If a~particle locates exactly on a~boundary between~$\hat V_k$ and~$\hat V_l$, then it is associated with the~subdomain~$\hat V_{\min\{k, l \}}$.

\begin{figure}[t]  
  \center\includegraphics[width=0.9\columnwidth]{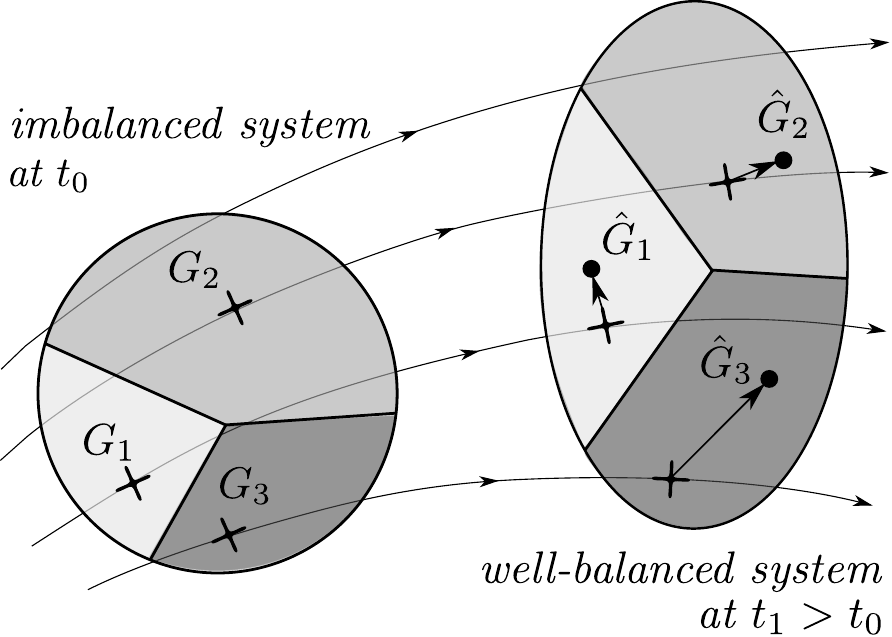}
  \caption{Displacement of the~Voronoi generator for each subdomain is determined by a~weighted sum of its material flow movement for a~simulation step and balancing displacements for pairs with surrounding Voronoi subdomains to equalize the~loads of the~corresponding processes.}
  \label{fig:Voronois}
\end{figure}

The \VD algorithm for SPH proposed in this paper allows a~generator to move following the~local material flow of the~SPH~particles assigned to the~corresponding subdomain, while the~balancing algorithm corrects the~displacements of generators to achieve a~load balance between subdomains  (i.e., between the~corresponding processes), as it was suggested in~\cite{zhakhovskii2005new} and is illustrated by Fig.~\eqref{fig:Voronois}. Once the~new boundaries for new generator positions are found, new assignment of particles is realized by sending all particles located beyond the~boundaries to the corresponding subdomains/processs using the~definition~\eqref{eq:VoronoiDefinition}.

The~information only about neighboring Voronoi subdomains, that are located within a~lookup radius~$R_{k} > |\vvec{g}_k - \vvec{g}_l|$, $l \ne k$, is required for~$\hat V_k$ to organize a~proper particle re-assignment. By exchanging particles with all adjacent subdomains/processes, the~$\hat V_k$ gets all particles located within its boundaries and gives away all that are not. Using the~distance~$R_{k}$ for determining the~subdomain neighborhood is necessary because a~mixing of subdomains is expected and Voronoi diagram connectivity may change. To reduce the~number of interprocess communications it is appropriate to choose~$R_{k}$ large enough to ensure that all neighbors are listed.

The~accuracy of simulation in a~large computational domain, where particle coordinates vary in several orders of magnitude, is subjected to truncation errors in the~calculation of interparticle distances. To reduce these errors, the~different local coordinate systems for each Voronoi subdomain~$\hat V_k$ are used instead of a~global computational domain system. A good choice for an origin of a~local coordinate system is a~geometrical center of~$N_k$ particles inside domain:
\begin{equation}\label{eq:VoronoiGeomCenters}
  \vvec{r}_{\text{c},k}=\frac{1}{N_{k}}\sum_{i=1}^{N_{k}}\vvec{r}_i.
\end{equation}
Then the~values of particle coordinates are restricted by subdomain dimensions. However, the~truncation errors may appear if this center moves with particles and also after particle transfer between different subdomains. To avoid the~effect of coordinate system change the~origins of local systems are assigned to the~nearest mesh site in which the~mesh size equals the~unit of length in use. Thus, the~coordinate of a~particle experiences occasionally an integer shift without truncation error, when the~origin obtains a~new mesh position or the~particle is transferred between subdomains. Generator coordinates are not rounded to the~mesh size, i.e., they remain in the~global computational domain system.

\begin{figure*}[t]
\begin{subfigure}[t]{0.24\linewidth}\centering
\includegraphics[height=1\textwidth]{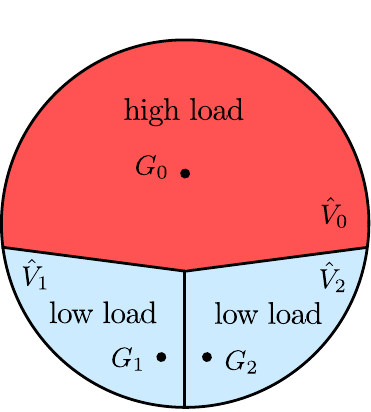}
\caption{$L_0 = L_h$, $L_{1,2} = L_l < L_h$}\label{loadrotateA}
\end{subfigure}
\begin{subfigure}[t]{0.24\linewidth}\centering
\includegraphics[height=1\textwidth]{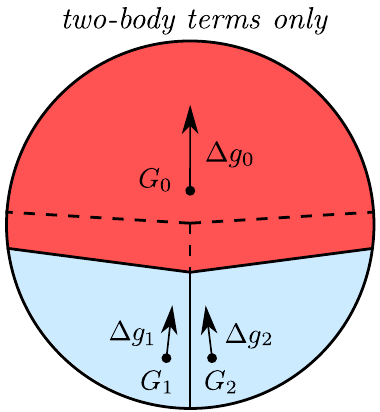}
\caption{$\Delta g_0\sim2(L_h-L_l)$, $\Delta g_{1,2}\sim(L_h-L_l)$}\label{loadrotateB}
\end{subfigure}
\begin{subfigure}[t]{0.24\linewidth}\centering
\includegraphics[height=1\textwidth]{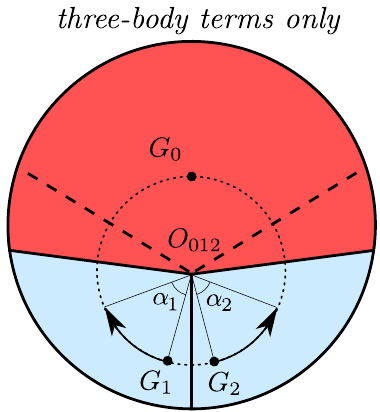}
\caption{$\alpha_{1,2}\sim (L_h-L_l)$, \\ $\alpha_0\sim0$}\label{loadrotateC}
\end{subfigure}
\begin{subfigure}[t]{0.24\linewidth}\centering
\includegraphics[height=1\textwidth]{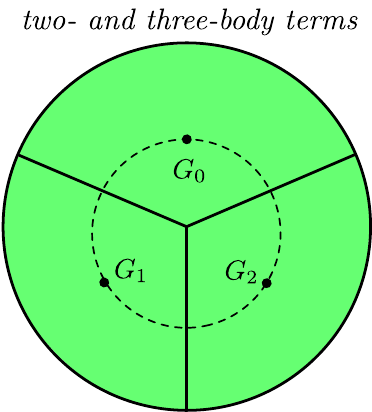}
\caption{$L_0=L_1=L_2$}\label{loadrotateD}
\end{subfigure}
\caption{Two-body and three-body balancing displacements for the simplest three subdomain decomposition of material disk. The~special initial decomposition (a) with severe imbalance is chosen to illustrate the failure of mere two-body displacements (b) to reach the best decomposition (d). Dashed lines indicate the final boundaries between Voronoi subdomains after many iterations. Pure rotation of the~generator points around the~common point $O_{012}$ of Voronoi subdomains leads to a~better balanced decomposition (c). Only combination of two- and three-body terms results in the best decomposition (d) with the minimal length of subdomain boundaries.}
\label{fig:loadRotate}
\end{figure*}

\subsection{Load balancing via two-body displacements of diagram generators}\label{subsec:Pair}
The~simplest load balancing formula appears as a~vector sum of two-body displacements which linearly depend on pairwise load imbalances between neighbor Voronoi subdomains \cite{zhakhovskii2005new}:
\begin{equation}
\label{eq:BalancingOffset}
  \Delta \vvec{g}_k=\sum_l^{M_k}\Delta \vvec{g}_{k, l}=\sum\limits_{l}^{M_k}\Dshell_{k,l}\frac{L_k-L_l}{L_k+L_l}\frac{\vvec{g}_{kl}}{|\vvec{g}_{kl}|},
\end{equation}
where~$\Dshell_{k,l}$ is a~half width of layer to send/receive particles introduced further in Section~\ref{subsec:SPHExchange},
$M_k$ is the~number of neighbor subdomains of~$\hat V_k$. The~length factor~$\Dshell_{k,l}$ is chosen empirically to provide the two-body displacements be within the thickness of boundary layer in order to keep the overhead time required for particle exchange between subdomains small enough.

This two-body algorithm~\eqref{eq:BalancingOffset} works very well if the local load imbalance is modest, because the sum of small two-body displacements gives a~good approximation of the unknown perfect balancing displacement in such conditions. Here this algorithm is improved for  the severe local imbalance by taking into account three-body terms in 2D decomposition and  four-body terms in 3D one.

\subsection{Three-body correction of balancing displacement for 2D Voronoi diagram}\label{subsec:Triple}

The usage of two-body load balancing displacements~\eqref{eq:BalancingOffset} may require many iterations in order to reach a~good local balance, if a~significant imbalance appears between the neighbor subdomains as a~result of the bad initial decomposition or abrupt changes of mass distribution in the simulated material flow. The~latter can be produced by shock compression, jet collisions, void formation and other high-rate processes which are typically produced by the extreme conditions. To speed up the convergence of balancing iterations a~higher approximation to the best balancing displacement depending on whole Voronoi diagram should be taken into account. Since the expression for such displacement is unknown we consider here the simplest decomposition to deduce a~form of three-body term which can improve the load balance for a~shorter time.

Let us consider a~2D decomposition of a~uniform material disk between three Voronoi subdomains as shown in~Fig.~\ref{fig:loadRotate}(a). Here the initial decomposition is chosen to gives the large difference $\Delta L = L_h - L_l \approx L_l\approx 0.25$ between the low load $L_l\approx 0.25$ and the high load $L_h\approx 2L_l\approx 0.5$, which are assumed to be proportional to the corresponding subdomain areas within the disk area of unity. It is clear that the usage of two-body formula~\eqref{eq:BalancingOffset} alone requires many iterations to reach a~good balance but will never lead to the perfect decomposition with the shortest subdomain boundaries shown in~Fig.~\ref{fig:loadRotate}(d). The~parallel displacements of all three subdomains are responsible for such inappropriate behavior.

To improve the balancing algorithm we suggest to use rotations around the common point $O_{012}$ of three contacting subdomains as shown in~Fig.~\ref{fig:loadRotate}(c). Such rotations are definitely required to push the subdomain positions in direction toward the best decomposition. To realize such rotations the three-body terms are added to the two-body balancing displacement~\eqref{eq:BalancingOffset} as follows
\begin{equation}
\label{eq:BalancingOffsetPlusRot}
  \Delta \vvec{g}_k=(1-\sigma)\sum_l\Delta \vvec{g}_{k, l} + \sigma\sum_l\sum_m\Delta \vvec{g}_{k, l, m},
\end{equation}
where the three-body terms are
\begin{multline}
\label{eq:BalancingOffsetRot}
\Delta \vvec{g}_{k, l, m} = -\vvec{c}_{k} +
\mathrm{M}\left(\vvec{c}_k \times \vvec{c}_{m}, \alpha_m \right) \mathrm{M}\left(\vvec{c}_{k}\times \vvec{c}_{l}, \alpha_l \right) \vvec{c}_{k},\\
  \vvec{c}_{p} = \vvec{g}_p - \vvec{o}_{klm}, \quad p = k,l,m.
\end{multline}
Here the~$\mathrm{M}(\vvec{a}, \alpha)$ matrix is a~rotation matrix which generates in-plane rotation by the angle~$\alpha$ around the~normal~$\vvec{a}$ at the common point $O_{klm}$.
The forms of angles $\alpha_l=\pi/3\cdot(L_l-L_k)/(L_k+L_l+L_m)$ and $\alpha_m=\pi/3(L_m-L_k)/(L_k+L_l+L_m)$ are chosen to provide the rotation resulting in the best angle of $120^\circ$ between the three subdomains shown in~Fig.~\ref{fig:loadRotate}d after single iteration.
The~parameter~$\sigma$ is a~weight of three-body terms. The~vector~$\vvec{o}_{klm}$ is a~radius-vector to position of $O_{klm}$, which is actually the~center of circumcircle touching the generator points~$G_k$, $G_l$, $G_m$. Such three-body correction in multi-subdomain decomposition can be applied for all triplets of contacting subdomains having the common point. To stabilize the balancing shift part $\sum_l\sum_m\Delta \vvec{g}_{k, l, m}$ we limit its length to $\max_{l=1,\ldots,M_k}\Dshell_{k,l}$ in dynamic simulations as a~change in the diagram connectivity can make the shift too large, which is undesirable as it causes extra useless particle exchanges.

The~relation~\eqref{eq:BalancingOffsetPlusRot} can be generalized to obtain four-body displacements~$\Delta \vvec{g}_{k, l, m, n}$ for balance correction in 3D diagram. The~3D rotation is performed on a~circumsphere of four generator points of contacting subdomains~$\hat V_k$, $\hat V_l$, $\hat V_m$, $\hat V_n$ having a~common point (normally four subdomains are in~contact in~3D diagrams). To illustrate the improvement of load balancing by
the multi-body terms we present only the tests of three-body correction for 2D diagrams in this work, see Section \ref{sec:tree-body-test}.

\subsection{The cumulative movement of Voronoi subdomain}
\label{subsec:TotalOffset}

A new position~$\hat{\vvec{g}}_k$ of~$k$-generator can be defined as a~weighted summation of average movement of all particles within~$\hat V_k$ during a~short period of time and the balancing displacement~$\Delta \vvec{g}_k$ as follows:
\begin{equation}
  \label{eq:TotalOffset}
  \hat{\vvec{g}}_k = (1 - \theta) (\vvec{g}_k + \gamma\Delta \vvec{g}_k) + \theta (\vvec{r}_{\text{c},k} + \Delta\vvec{r}_{\text{c},k}),
\end{equation}
where~$\vvec{g}_k$ is the previous position of generator and $\vvec{r}_{\text{c},k}$ the previous geometrical center of the subdomain~\eqref{eq:VoronoiGeomCenters}, $\Delta\vvec{r}_{\text{c},k}$ is a~current average Lagrangian displacement vector of those particles, and~$\theta \in [0,1]$ is a~parameter specifying whether the~Voronoi subdomain movement is preferably determined by the~movement of particles~\eqref{eq:VoronoiGeomCenters}
or by the~load balancing mechanism~\eqref{eq:BalancingOffsetPlusRot}. The~adding of the geometrical center of all particles $\vvec{r}_{\text{c},k}$ in the cumulative Voronoi subdomain movement~\eqref{eq:TotalOffset} makes it responsive to any particle exchange between subdomains even in static samples with material at rest. It drives the subdomain shapes to more regular forms, which minimizes the total length of subdomain boundaries. As a~result the balance convergence is accelerated and the overall parallel performance is improved.

A simplified diagram shown in~Fig.~\ref{fig:Voronois} illustrates the linear combination of the balancing displacement, material movement and redistribution between subdomains. Static spatial decomposition is realized by~$\gamma = \theta = 0$. If~$\theta=1$ there is no balancing displacements and the~evolution of \VD is exclusively controlled by the~material motion. Theoretically, the~choice of~$\theta = 0$ provides the maximal convergence rate but it is inadequate for simulation of fast material motion as it leads to irregular subdomain shapes and large jitter of their positions, which  results in extensive data exchange overheads. The~parameter $\theta$ should be chosen empirically to provide a~shorter simulation time on a~specific cluster, because it depends on the bandwidth of internode fabric in use. The~$\theta = 0.25$ is used in test modelings that we discuss below in~Sections \ref{sec:static_results} and~\ref{sec:dynamic_results}.

The~presented load balancing strategy results in that a~heavy-loaded worker attracts less-loaded neighbors and distributes its excess particles among them. The~process is running iteratively, reducing the~communication waiting time and achieving the~better parallel efficiency step by step. 
To reduce the~computational overheads for re-balancing at each simulation step, the~load balancing algorithm is executed once per a~time interval consisting of~$\Nupd$ simulation steps in our code. During this interval, the~SPH~particles are not allowed to be reassigned between subdomains. 

\begin{figure}[t]
\centering
    \begin{subfigure}[b]{0.45\linewidth}\centering
        \includegraphics[width=0.95\textwidth]{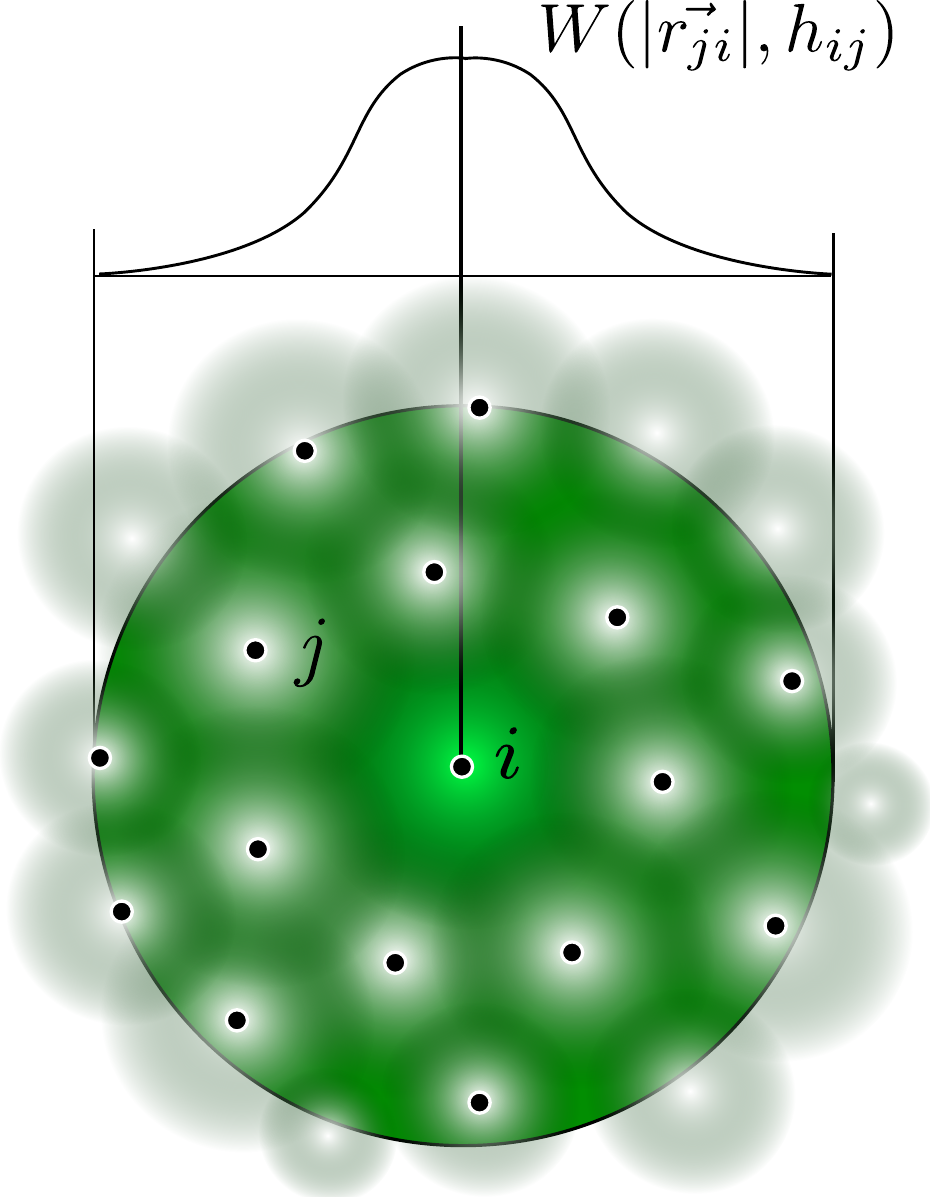}
        \vfill
        \caption{}\label{fig:SPHscheme}
    \end{subfigure}~
    \begin{subfigure}[b]{0.55\linewidth}\centering
        \includegraphics[width=\textwidth]{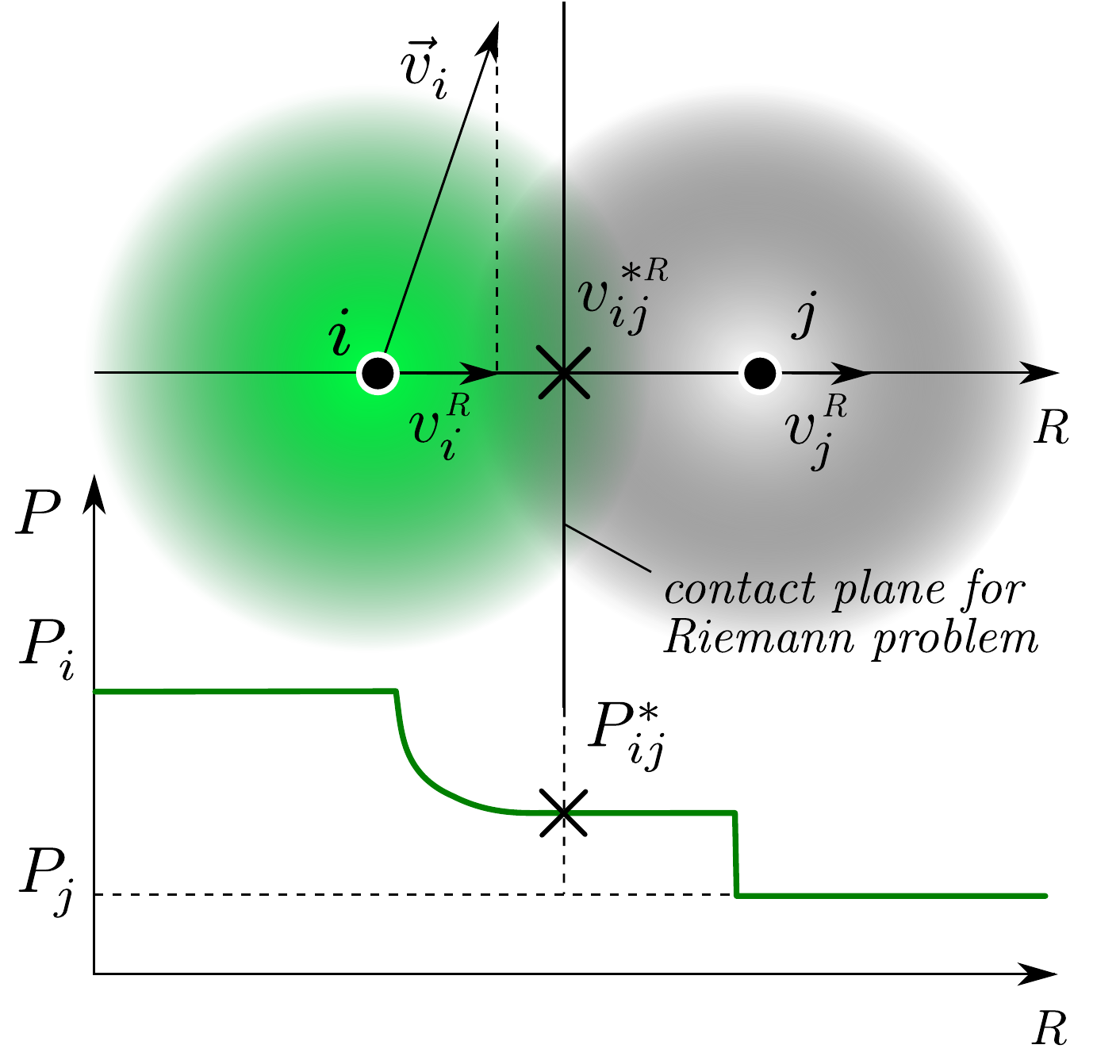}
        \caption{}\label{fig:CSPH}
    \end{subfigure}
    \caption{(a) Interaction of the~central particle~$i$ with neighboring particles~$j$ in the SPH method. (b) SPH with the incorporated Riemann problem solution. Interparticle interaction is approximated with Riemann problem solution at the interface between a~pair of particles.}
    \label{fig:SPH}
\end{figure}

\begin{figure}[t]  
\center\includegraphics[width=1.\columnwidth]{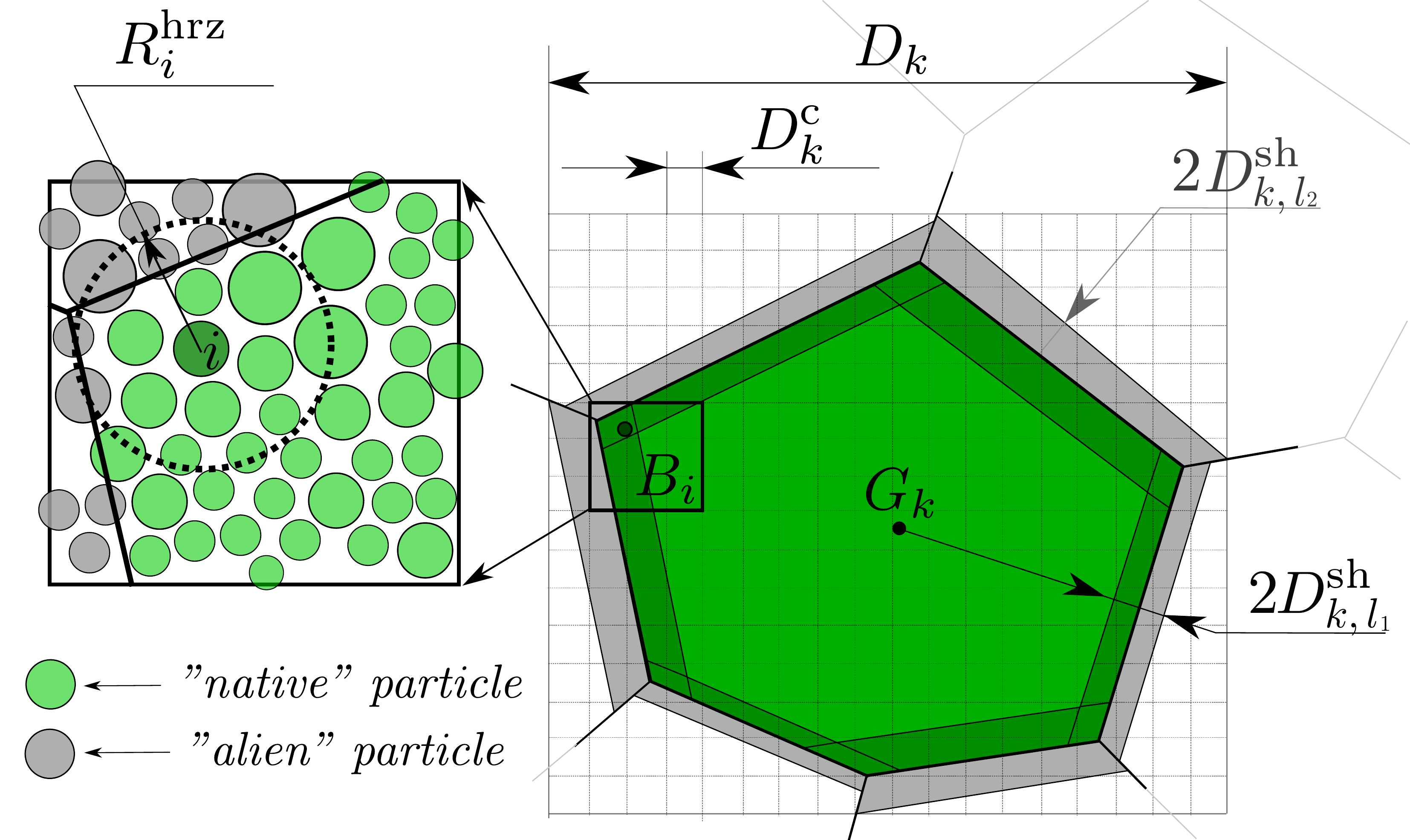}
\caption{A uniform cubic lattice with cell size $\Dcell$ is constructed in a~Voronoi subdomain corresponding to the generator point $G_k$. The~green-colored Voronoi subdomain is surrounded by the gray zone that belongs to the neighbor subdomains. Zone width and alien particles within are synchronized with the neighbor subdomains at a~neighbor list update. The~darker green area is a~part of Voronoi subdomain $V_k$, information about which the process $k$ provides to its neighbors. The~area $B_i$ of the linked list for the SPH particle $i$ is shown in the left inset.}
\label{fig:Voronoi}
\end{figure}

\section{Application of \VD to contact SPH}\label{sec:SPH-VD3}
\subsection{Smoothed particle hydrodynamics with incorporated interparticle Riemann problem solution}\label{sec:particlemethods}

The~Voronoi auto-balancing decomposition algorithm is suitable for any numerical method that represents material as a~set of discrete objects interacting by short-range forces. In SPH, each particle interacts only with those particles that are within the~smoothing kernel~$W(|\vvec{r}_{ij}|,h_{ij})$ of support area, where $h_{ij}$ is a~smoothing length between the~particles $i$ and $j$ (see Fig.~\ref{fig:SPHscheme}). The~limited interaction radius makes it possible to take into account the~interaction of SPH particles only within the~Voronoi cell and with other cells' bordering particles in the~zone determined by the~radius of the~smoothing kernel. In this case, load balancing can be efficiently performed by particle transfer between adjacent Voronoi cells.

Interaction of particles in CSPH \cite{parshikov2002smoothed,zhang2013contact} is provided by the~Riemann solver applied for each pair $i$--$j$ of interacting particles $i$ and $j$, as illustrated by Fig.~\ref{fig:CSPH}. The~velocity components $v_i^R =\vvec{v}_i \cdot \vvec{e}^ R$, $v_j ^ R = \vvec{v}_j \cdot \vvec{e}^ R$ of the~particles $i$ and $j$ are used as left and right state velocities, where $\vvec{e}^ R = ({\vvec{r}_j- \vvec{r }_{i}}) / {| \vvec{r}_j- \vvec{r}_{i}|}$ is a~direction vector. Then, the~velocity $\upsilon^{*R}_{ij}$ and pressure $P_{ij}^*$ can be obtained by the~Riemann solver. CSPH-approximations of conservation laws can be found as:
\begin{equation}\label{SPH:Conti}
\frac{d\rho_i}{dt}={2}{\rho_i}\sum_j\frac{m_j}{\rho_j}(v_i^R - v_{ij}^{*R}) \vvec{e}^R\cdot\nabla_i W_{ij}
\end{equation}
\begin{equation}\label{SPH:Pulse}
\frac{d\vvec{v}_i}{dt}=-\frac{2}{\rho_i}\sum_j\frac{m_j}{\rho_j}P^{*}_{ij} \nabla_i W_{ij}.
\end{equation}
Here $r_{ij} = |\vvec{r}_i - \vvec{r}_j|$, $W_{ij}=W(r_{ij}/h_{ij})$ is a~smoothing kernel between particles $i$ and $j$, $h_{ij}$ is a~smoothing length. The~kernel gradient equals to $\nabla_i W_{ij} = -\vvec{e}^R W'_{ij}$. Unlike~\cite{parshikov2002smoothed}, our code uses a~conservative formulation of the~energy conservation law:
\begin{equation}\label{SPH:Energy}
\frac{dE_i}{dt}=-\frac{2}{\rho_i}\sum\limits_j\frac{m_j}{\rho_j}P_{ij}^*v_{ij}^{*R} \vvec{e}^R\cdot \nabla_i W_{ij},
\end{equation}
where the~total energy $E_i = e_i + v_i^ 2/2$ is the~sum of the~internal and kinetic energies of a~unit mass in $i$-particle.

The~high energy density physics usually addresses the~problems where the~SPH particles can contract or expand easily, so the~variability of the~smoothing length is significant. The~particle size is defined as $d_i=\sqrt[3]{{m_i}/{\rho_i}}$. To ensure the~pair symmetry of interparticle interaction, it is necessary to specify the~symmetric smoothing distance:
\begin{equation}\label{EQ:SmoothingLengthDef}
h_{ij} = \left(d_i+d_j \right) / 2.
\end{equation}
The~interaction distance of the~particles depends on a~chosen smoothing kernel, and generally $W(r_{ij} / h_{ij})> 0$ for $|\vvec{r}_{ji}| < \kappa~h_{ij}$, where $\kappa$ is a~radius of smoothing kernel. All results presented in Sections~\ref{sec:static_results} and~\ref{sec:dynamic_results} are obtained using the~Wendland kernel $\mathrm C^2$~\cite{dehnen2012improving} with the~radius of $\kappa~= 1.936$.

\subsection{Neighbor particle lists used in Voronoi subdomains}\label{subsec:SPHNeighbours}
For the direct search of neighboring particles, the $\mathcal{O}(N)$ operations are required to find all particles $j$ interacting with a~chosen central particle $i\in [1,N]$. The~search for all interacting pairs requires $\mathcal{O}(N^2)$ operations. Given the finite interaction radius, the number can be reduced to $\mathcal{O}(N)$ using a~list of neighbor particles.

Generation of a~neighbor list for particles within each Voronoi subdomain takes a~considerable amount of time. Moreover, the neighbor list which includes ``alien'' particles from adjacent boundary layers with each surrounding Voronoi subdomains is required to find all neighbor particles. To reduce the computational overheads for neighbor list generation the list can be updated once in several time steps by utilizing the Verlet approach~\cite{allen1989computer}. The~idea is that neighbors of an individual particle are considered within a~sphere which includes not only interacting particles but also ones in a~buffer shell with a~large enough horizon radius $\Rhrz$ which is larger than an interaction radius. During several simulation steps this buffer may provide particles for the further interaction without update of the neighbor list. Such an enhanced neighbor list is called as the Verlet list (VL). The~``alien'' particles are also included in the VL for several steps $\Nupd$ between the list updates. The~list of neighbors is updated after each rebalancing of Voronoi decomposition, when boundary layers may change and the transfer of particles between the processes is allowed.

The efficiency of using the VL depends much on the buffer size and number of simulations steps which may proceed without the list update. In compressible SPH it is convenient to consider neighbors of a~particle $i$ in the individual sphere of the horizon radius $\Rhrz_{i} = \Rint_i + \Rbuf_i,$ where $\Rint_i$ is a~maximal local interaction distance, $\Rbuf_i = \beta\Rint_{i}$ is a~buffer size with the parameter $\beta>0$. If the buffer size is too small ($\beta \to 0$) the neighbor list must be constructed at each simulation step to be sure that interparticle interaction is calculated properly. The~number of neighbors within $\Rhrz$ grows as $~(1 + \beta)^3$, thus a~larger $\beta$ results in greater memory consumption and necessity to check distances with very many particles at each step that also increases computing time. Then an optimal $\beta$ and a~number of steps between updates $\Nupd$ can be found in a~few trial tests. We use $\Nupd = 10 \div 15$ and $ \beta = 0.3\div 0.6$ depending on specific simulation scenario.

To avoid calculation of $\mathcal{O}(N_k^2)$ iterparticle distances in order to construct the VL, all particles are associated with cubic cells of a~mesh, which covers a~Voronoi subdomain together with its ``alien'' particles in surrounding boundary layers as shown on Fig.~\ref{fig:Voronoi}. The~two biggest particles with diameters $d^{\max}_1$ and $d^{\max}_2$ should be found in the Voronoi subdomain and boundary layers to define the cell size $\Dcell$
\begin{equation}\label{eq:Dcell}
  \Dcell= \frac12 (1+\beta) \kappa (d^{\max}_1+d^{\max}_2).
\end{equation}
where $\kappa$ is determined in section~\ref{sec:particlemethods}. Such cell size definition guarantees that all ``native'' and ``alien'' neighbors of a~central particle are allocated within $3^3=27$ cells around it. This region is denoted as $B_i$ in~Fig.~\ref{fig:Voronoi}.

\algrenewcommand{\algorithmiccomment}[1]{\hspace{2 cm}$\triangleright$ {\it  #1}}
\begin{algorithm*}
  \caption{Data exchange and decomposition update after $\Nupd$ simulation steps}
  \label{algo}
  \begin{algorithmic}[1]
  \Procedure{}{}\\
  \Comment {/*Update Voronoi decomposition for all $k$-processes*/}
  \State Make balancing shift $G_k$ using Eq.~\eqref{eq:TotalOffset} from subsection~\ref{sec:Voronoi}.
  \State Reassign particles between $\hat V_k$ and its neighbors $\hat V_l$ using definition~\eqref{eq:VoronoiDefinition} from subsection~\ref{sec:Voronoi}.
  \State Calculate exchange layer width $\Dshell_{k, l}$ from Eq.~\eqref{EQ:Horizon} from subsection~\ref{subsec:SPHExchange}.
  \State Send lists of ``native'' particles in the exchange boundary layers to~$\hat V_l$.
  \State Obtain lists of ``alien'' particles in the exchange layers from neighbors $\hat V_l$.
  \State Build two separate Verlet lists for ``native'' and ``alien'' particles in $\hat V_k$. \\
  \Comment {/*Perform next $\Nupd$ simulation steps*/}
  \For{$i=1$ to $\Nupd$}
    \State request for information about ``alien'' particles;
    \State do SPH calculations of interaction between ``native'' particles and ``natives'';
    \State get information about ``alien'' particles;
    \State do SPH calculations of interaction between ``native'' particles with ``aliens'';
    \State renew particle positions and states and make global synchronization of SPH time step;
    \State count the useful time $\tu_k$;
  \EndFor
    \State Count the elapsed time $\te_k$ to evaluate a~new load $L_k,$ and repeat this procedure.
  \EndProcedure
  \end{algorithmic}
\end{algorithm*}

Indices of particles within all cells of the mesh are stored in a~linked list (LL) having a~form of integer vector, which size is equal to number of all $N_k$ particles within a~given Voronoi subdomain. To keep entry indices for LL a~3D integer array with dimensions corresponding to the mesh is allocated. If a~cell is not occupied by particles the corresponding element of this 3D array is marked as zero. Otherwise, the index of first particle found in each cell is stored in the 3D array. This index points to the address of LL element where a~next particle index in the same cell is allocated, as it is described in ~\cite{allen1989computer}. This linked procedure continues until the particle index points to a~LL element containing zero, which ends the particle list for the cell. Using the linked list, the Verlet list can be generated performing $\mathcal{O}(N_k)$ calculations of distances between particles located only in surrounding $3^3$ cells.

A list of particles within $B_i$ obtained from the LL is used to determine the maximal interaction distance $\Rint_i$ for $i$-particle as follows:
\begin{equation}
  \Rint_i = \frac12 \kappa \left( d_i + \max_{j\ne i \in B_i}d_j\right),
\end{equation}
and then the horizon radius is given:
\begin{equation}
  \Rhrz_i = (1 + \beta)\Rint_i,
\end{equation}
which is definitely smaller than $\Dcell$ from Eq.~\eqref{eq:Dcell}. Such extended lists of neighbor particles for all $B_i$ are used to build the VL containing only neighbors indices of particles which are placed within $\Rhrz_i$.

Since modern computing systems have several levels of the memory hierarchy (from fast low-level CPU cache to slow RAM), simulation can be accelerated by increasing the probability of cache hits. To do so the particles, which spatially close to each other in simulation domain, should have the closer addresses in a~linear RAM space. Then the calculation of their interactions is likely to be executed with using particle coordinates allocated in fast CPU cache. To utilize this idea the re-numbering and memory reallocation of neighbor particles using the Verlet list is performed from time to time. This procedure helps to decrease the cache missing rate, which is specifically amplified by extensive particle exchange between Voronoi subdomains, at which the arrived particles got new indices leading to the wide spread of particle indices in the Verlet lists. If the number of particles per a core is not too small and the particle exchange involves a significant number of particles at each step, the re-numbering may provide a speedup up to 2 times with respect to a simulation in which such memory re-ordering of particles is not used.

\subsection{Adjusting interprocess data exchange for SPH}\label{subsec:SPHExchange}

It is clear that the minimal width of boundary layer should be chosen to include all interacting particles near the boundary between Voronoi subdomains. As soon as we use the Verlet neighbor list it is necessary to acquire information about ``alien'' particles within a~surrounding lookup area at every time step. Then for all pairs of contacting Voronoi subdomain $k$ and $l$ the width of boundary layer $\Dshell_{k,l}$ must be determined. Using the maximal horizon radii in the Voronoi subdomains $\hat V_k$ and $\hat V_l$ we determine a~layer width $\Dshell_{k,l}$ in a~symmetric pairwise form as follows:
\begin{equation}\label{EQ:Horizon}
\Dshell_{k,l} = \max \left[
\max_{i\in \hat V_k} \Rhrz_i, \,
\max_{i\in \hat V_l} \Rhrz_i
\right].
\end{equation}
That definition guarantees that the buffer zone with horizon radius $\Rhrz_i$ for each particle near the boundary is correctly filled with the ``alien'' particles. It is required to generate the identical lists of interactive ``native--alien'' particle pairs in both contacting Voronoi subdomains.

The boundary layer width~\eqref{EQ:Horizon} is then directly linked to the particles buffer size $\Rbuf$. By varying parameters $\beta$ and $\Nupd$, one may get not only optimal neighbor list length but also minimize the amount of data transfers and reduce both the exchange and elapsed times. It is possible to develop an algorithm for dynamic adjustment for $\beta$ and $\Nupd$ during simulation aiming to reduce the elapsed time depending on simulation conditions, but this question is beyond the scope of this paper.

It should be noted that each process associated with Voronoi subdomain does not calculate ``alien--alien'' interaction. It receives the updated particles lists from its subdomain neighbors at each simulation step and handles only ``native--native'' and ``native--alien'' interactions. The~code for these two parts can be executed separately, that provides the opportunity to mask waiting for ``alien'' particles by using asynchronous non-blocking MPI-communications and get them during calculation of ``native--native''.

The pseudocode listed in Algorithm \ref{algo} summarizes all parts of the \VD and the auto-balancing algorithm described in the above subsections of this section~\ref{sec:AllVoronoi}.

\begin{figure*}[t]
\begin{subfigure}[t]{0.195\linewidth}
\includegraphics[width=1\linewidth]{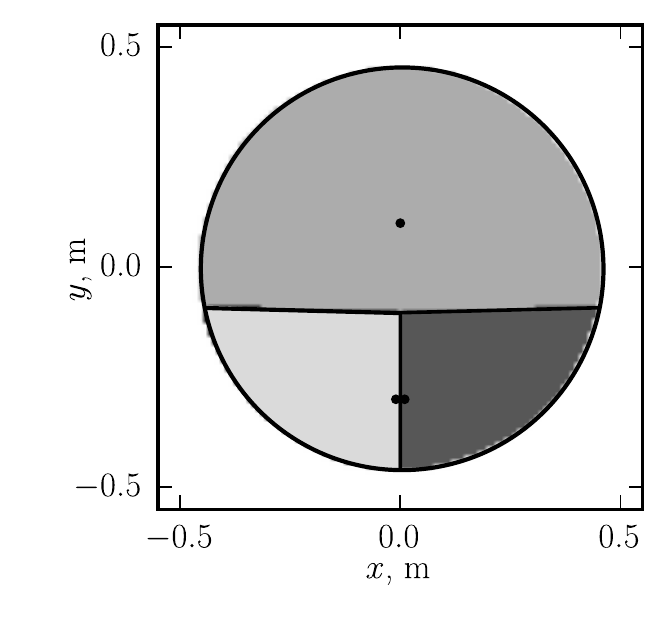}
\caption{Initial state}\label{fig:loadRotate-a}
\end{subfigure}
\begin{subfigure}[t]{0.195\linewidth}
\includegraphics[width=1\linewidth]{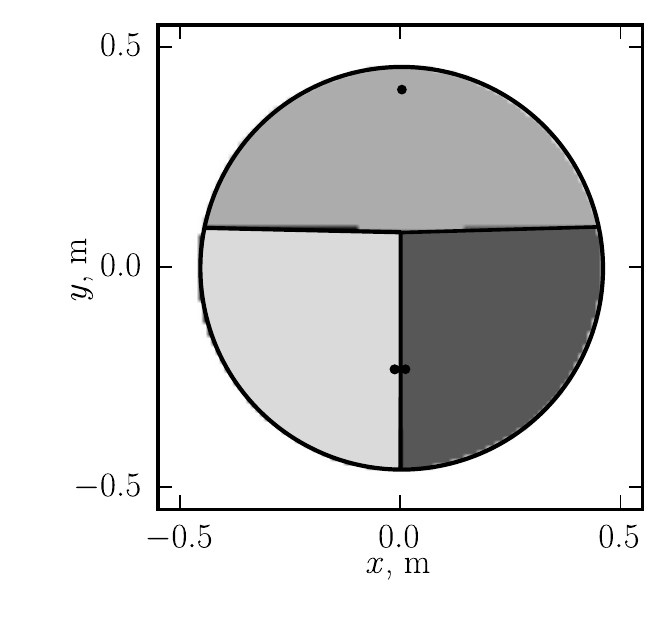}
\caption{{$\sigma = 0$, $\theta=0$, }\\{21 VD$^3$ iterations}}\label{fig:loadRotate-b}
\end{subfigure}
\begin{subfigure}[t]{0.195\linewidth}
\includegraphics[width=\linewidth]{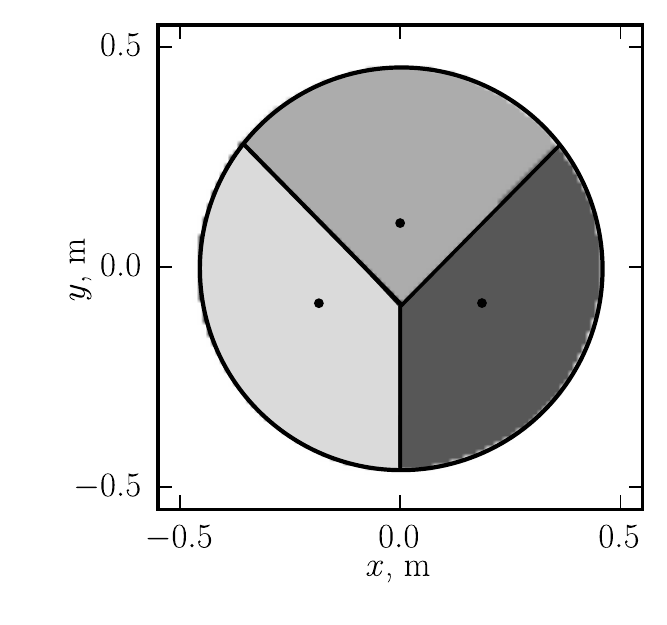}
\caption{{$\sigma = 1$, $\theta=0$, }\\{21 VD$^3$ iterations}}\label{fig:loadRotate-c}
\end{subfigure}
\begin{subfigure}[t]{0.195\linewidth}
\includegraphics[width=\linewidth]{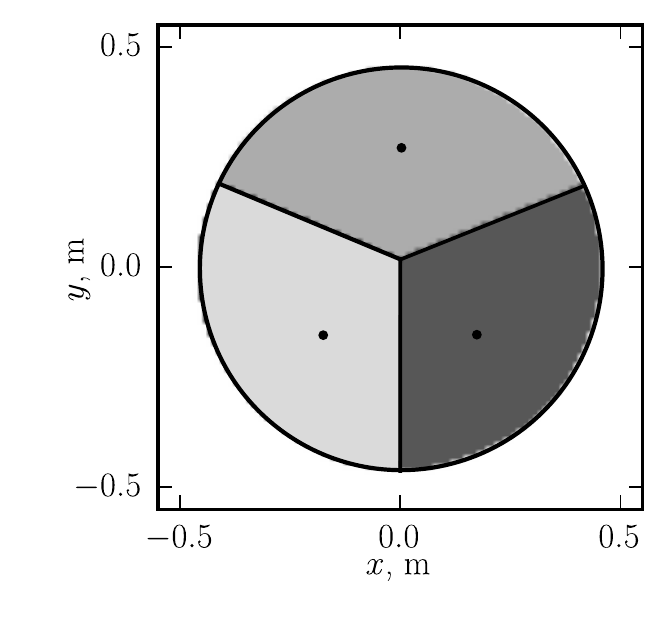}
\caption{{$\sigma=0.5$, $\theta=0$, }\\{17 VD$^3$ iterations}}\label{fig:loadRotate-d}
\end{subfigure}
\begin{subfigure}[t]{0.195\linewidth}
\includegraphics[width=\linewidth]{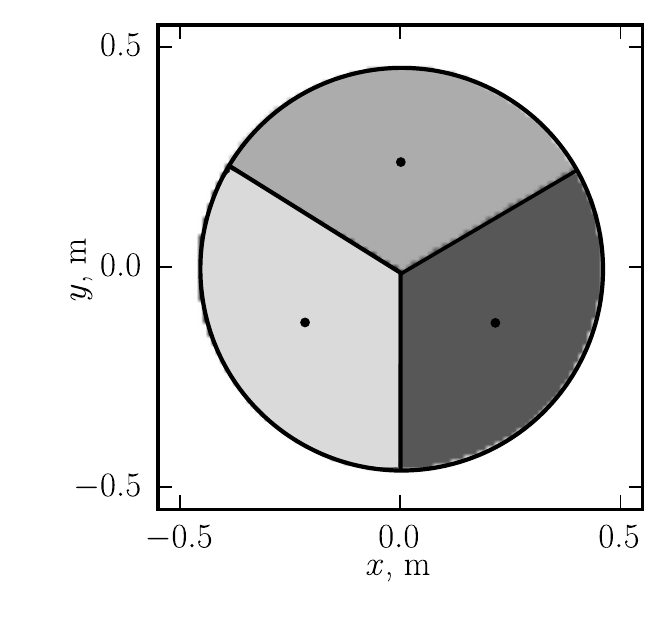}
\caption{{$\sigma=0.5$, $\theta=0.25$, }\\{11 VD$^3$ iterations}}\label{fig:loadRotate-e}
\end{subfigure}

\caption{Decompositions attainable from an initial bad one (a) by the different balancing algorithms using: (b) the two-body terms only, (c) the three-body terms only, (d) the combined two- and three-body balancing displacements, and (e) the cumulative Voronoi movements by Eq.~\eqref{eq:TotalOffset}, which leads to the perfect decomposition. Simulations were performed with material disk at rest. The~shown frames correspond to the iteration numbers when the convergence condition~\eqref{eq:VDConvergence} is satisfied for the first time. The~corresponding load balances as functions of iteration number are presented in~Fig.~\ref{fig:Convergence}. }
\label{fig:loadRotate}
\end{figure*}

\begin{figure}[t]
\center
\includegraphics[width=0.7\linewidth]{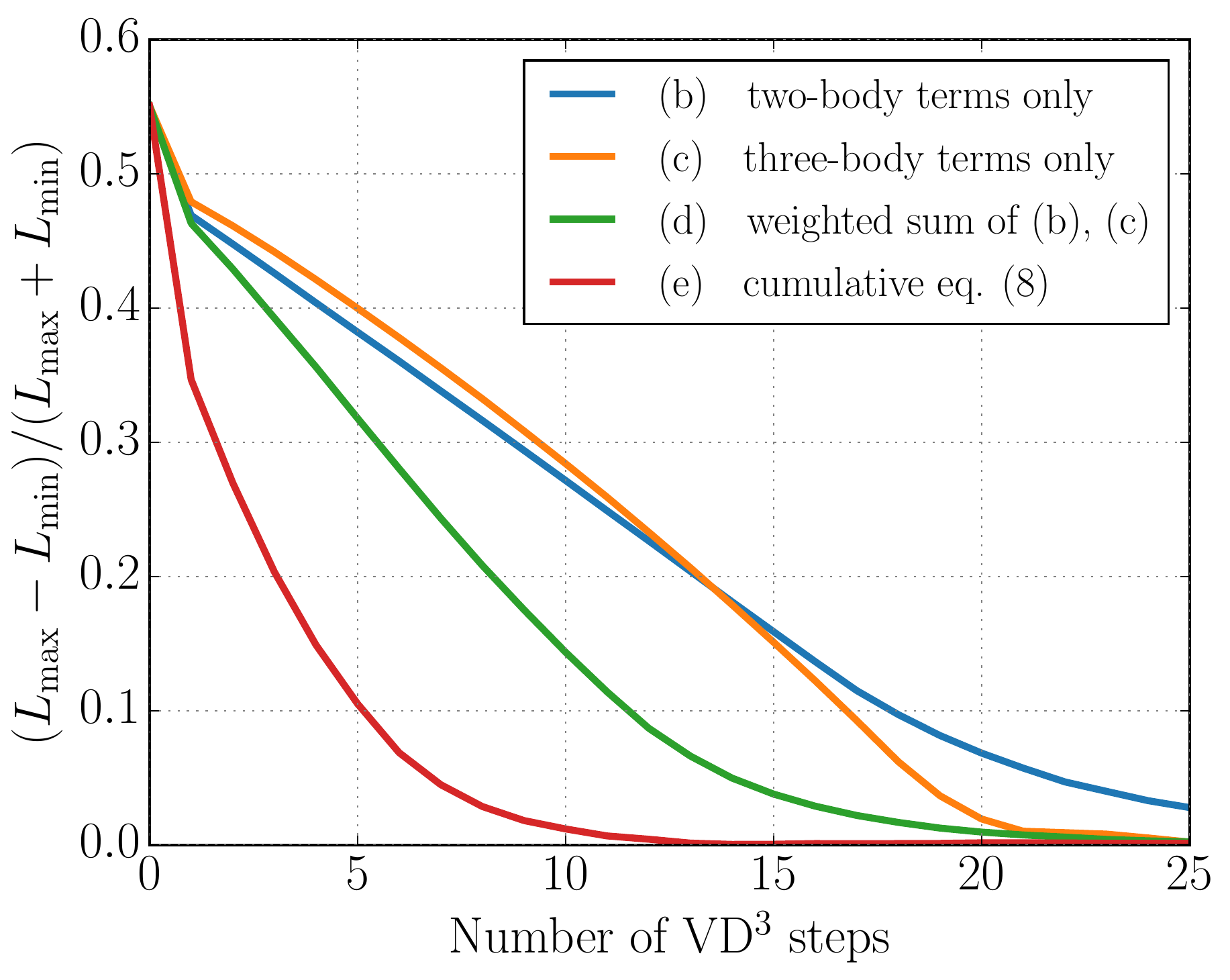}
\caption{Convergence of different balancing algorithms starting from an imbalanced decomposition shown in~Fig.~\ref{fig:loadRotate-a}. A well-balanced decomposition is not established for the mere two-body displacements, see the corresponding frames in~Fig.~\ref{fig:loadRotate}(b--e). The~fastest convergence is realized by the cumulative movement formula \eqref{eq:TotalOffset}.}
\label{fig:Convergence}
\end{figure}

\section{Convergence of balancing algorithm with different displacements}
\label{sec:tree-body-test}

The cumulative movement of Voronoi subdomain \eqref{eq:TotalOffset} depends on the material motion, the shift of geometrical center due to particle exchange between the subdomains, and the balancing displacement. The~latter consists of two-body and three-body terms \eqref{eq:BalancingOffsetPlusRot}. The~convergence with using these terms is tested by simulations of a~simple system similar to that discussed above in subsection \ref{subsec:Triple}.

A disk of material at rest with the height of~$0.1$~m and radius of~$0.45$~m is represented by uniformly distributed 127\,230 SPH particles. Initially the disk is decomposed between three subdomains as shown in Fig.~\ref{fig:loadRotate-a}. Then this imbalanced decomposition is used as a~starting point in four convergence tests to obtain the maximal difference of loads as a~function of iteration number. Figure~\ref{fig:loadRotate} shows the frames with decomposition at the iteration numbers for which the following convergence condition is met for the first time:
\begin{equation}\label{eq:VDConvergence}
\sum_{k=1}^3 |\hat{\vvec{g}}_k - \vvec{g}_k|<0.01.
\end{equation}

The convergence of the load difference obtained by the different balancing algorithms is presented in~Fig.~\ref{fig:Convergence}. The~goal is to check convergence to the best decomposition in given test, which consists of three equal subdomains having the common point at the disk center. Such decomposition has also the minimal length of subdomain boundaries which provides the smallest data exchange between the corresponding processes.

The mere two-body displacements are realized by the parameters $\sigma=\theta=0$ and $\gamma = 1$ in~Eqs.~\eqref{eq:BalancingOffsetPlusRot} and \eqref{eq:TotalOffset}. Such algorithm cannot change the angles between subdomains as their movements are oriented to the single heavy loaded subdomain, which preserves the initial decomposition geometry as seen in~Fig.~\ref{fig:loadRotate-b}. As a~result, this two-body algorithm gives the slowest convergence and fails to reach the best decomposition in the given test.

By contrast, the mere three-body displacements (at $\sigma=1$, $\theta=0$ and $\gamma = 1$) are able to rotate the subdomains and reach the better angle between them as shown in~Fig.~\ref{fig:loadRotate-c}. However this algorithm cannot move a~common point of three contacting subdomains. It leads to a~slow convergence and the best decomposition cannot be reached again.

The combination of two- and three-body terms ($\sigma=0.5$, $\theta=0$ and $\gamma = 1$) accelerates the convergence noticeably as seen in~Fig.~\ref{fig:Convergence}. Such fast convergence can be even faster if the balancing algorithm utilizes Eq.~\eqref{eq:TotalOffset} for cumulative movement of Voronoi diagram is applied. It leads to the best decomposition shown in~Fig.~\ref{fig:loadRotate-e} for several iterations.

\section{Parallel performance in static tests}\label{sec:static_results}

The primary goal of the \VD with load balancing algorithm is to reduce a~wall clock time required for simulation of nonuniform flow of materials. But first we have to check the strong scalability of \VD in static tests where material stays at rest all the simulation time, and only the number of CPU cores and positions of Voronoi generators can change. That allows testing the strong scalability in ideal conditions, without an effect of material motion on the realignment of decomposition. Dynamic testing the adaptive behavior of the algorithm under conditions of high-rate mass redistribution is discussed in the next Section~\ref{sec:dynamic_results}.

All static and dynamical tests were performed on our institute-wide 96 node cluster connected by InfiniBand (4X FDR) fabric with 56 Gb/s bandwidth. Each node is equipped by two 8-cores CPU Xeon E5-2670 operated at 2.6 GHz.

The static tests with a~steady sample are performed to trace the rearrangement of the initial decomposition to a~well-balance one using the balancing algorithm and the dependence of calculation speedup on a~number of processes involved. 2D decomposition is applied to a~quasi-two-dimensional sample for clarity of decomposition visualization. A thin square plate of material with sizes $L_x=L_y=1$~m, $L_z=0.015$~m is considered at rest. The~periodical boundary conditions (PBC) are imposed along all axes. The~material is represented by about 52 millions of SPH particles. The~number of particles remains constant in all static tests in order to check the strong scalability of our code, that is, the growth of calculations speedup with the increase of CPU cores $N_p$ involved in decomposition.

\begin{figure*}[t]    
\center\includegraphics[width=0.44\textwidth]{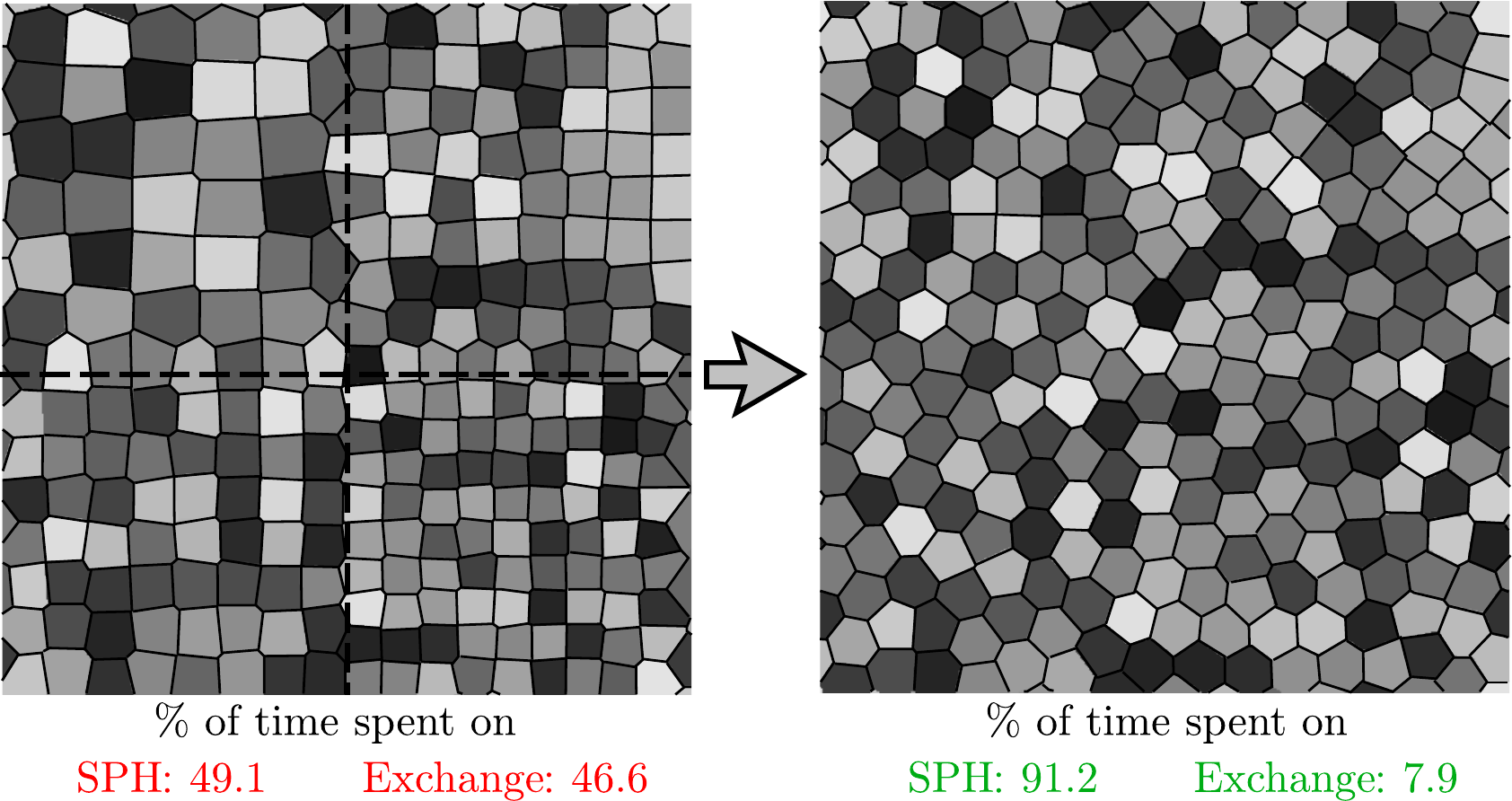}
\includegraphics[width=0.26\textwidth]{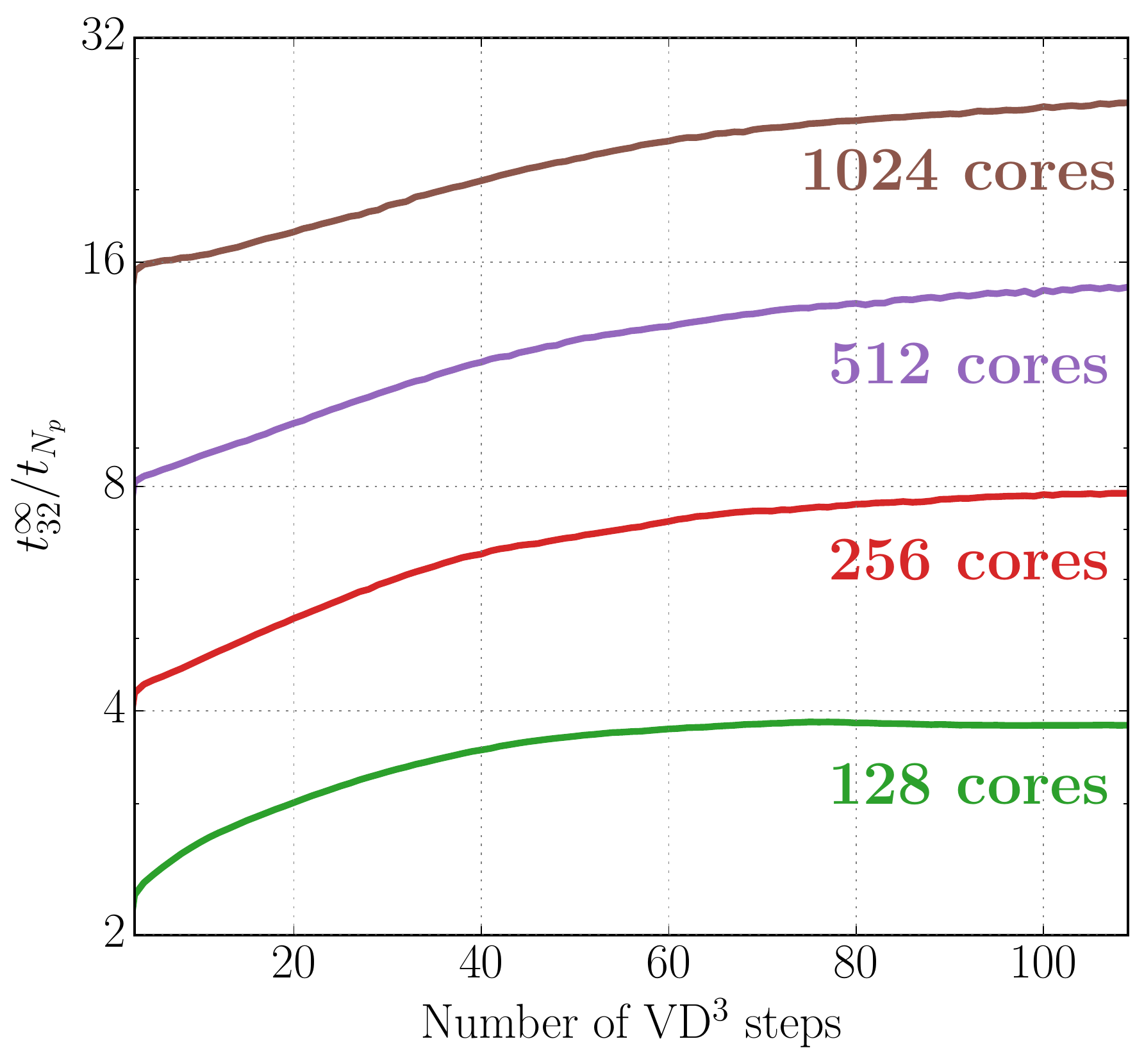}
\includegraphics[width=0.26\textwidth]{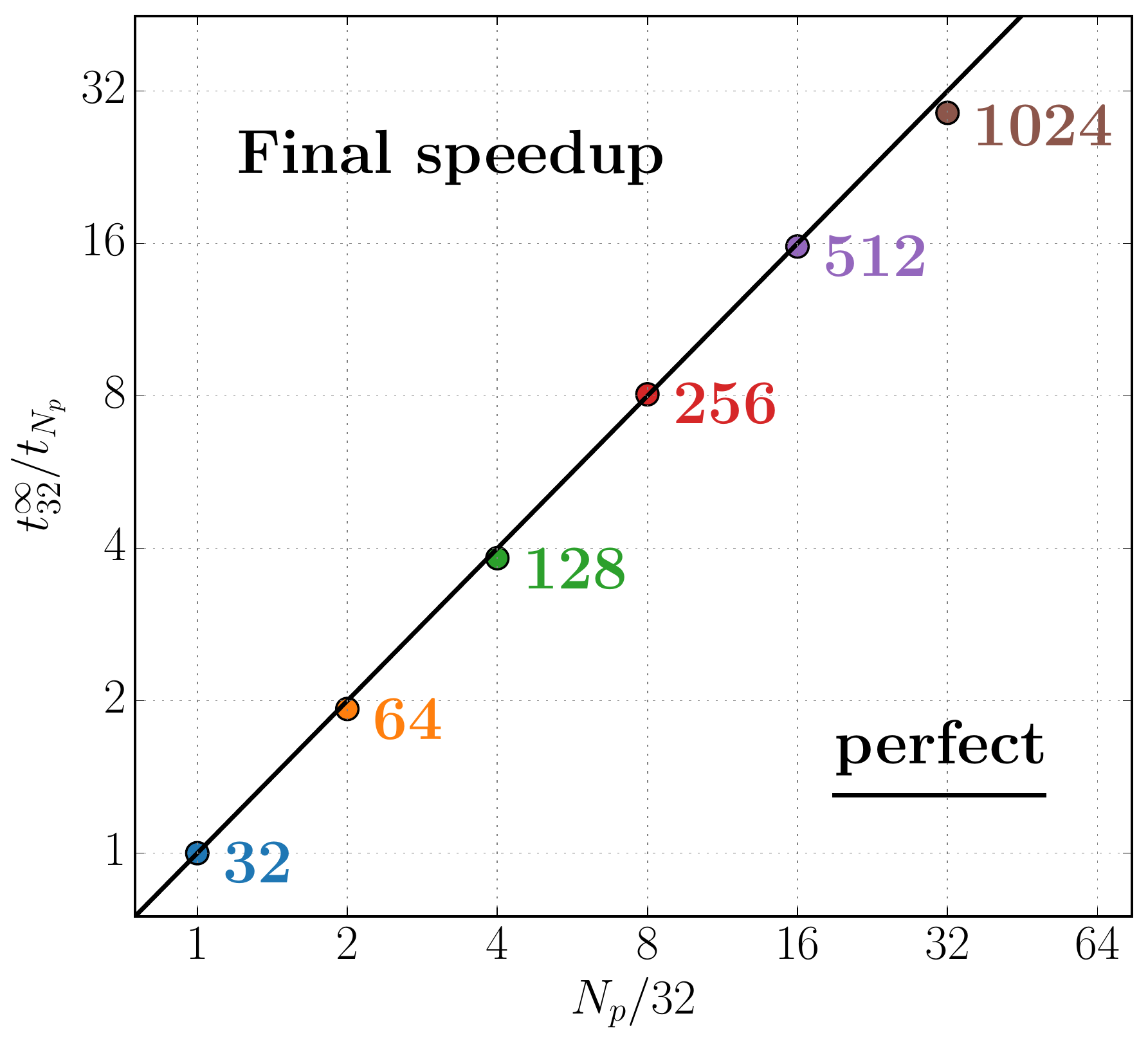}\\
\hspace{0.1\linewidth}(a)\hspace{0.22\linewidth}(b)\hfill(c)\hfill\qquad(d)\hspace{0.12\linewidth}\null \\
\caption{Stationary thin plate of sizes $L_x = L_y = 1$~m, $L_z = 0.15$~m consisting of 52 millions of particles is used for static tests of \VD and autobalancing algorithm. Tests run on 32, 64 128, 256, 512, 1024 CPU cores with initial imbalance introduced into system. Fig.~(a) shows initial imbalanced Voronoi diagram, where one quarter of the sample consists of $N_p / 8$ Voronoi subdomains, one consists of $3N_p/8$ and two---of $N_p/4$. Transition from initial to well-balanced decomposition with 256 cores is demonstrated on the Figs.~(a) and (b). Increase of speedup with iterations in transition from unbalanced to balanced decomposition is shown in the Fig.~(c). Elapsed time per a~step in well-balanced system is denoted as $t^{\infty}$. Final speedup as a~function of CPU core number is almost linear for well-balanced decomposition as shown in the Fig.~(d). }
\label{fig:static}
\end{figure*}

Initially the load imbalance is introduced to the system by splitting each quarter of square sample by Voronoi subdomains in proportions $3N_p/8$, $N_p/4$, $N_p/4$, $N_p/8$ of total $N_p$ MPI-processes as shown on Fig.~\ref{fig:static}a. To initiate the decomposition rearrangement inside those quarters, small random shifts to initial Voronoi generator positions were applied. Due to memory limitation in a~computing node of the used computing cluster, $N_p$ is only varied from 32 to 1024.

Initial imbalance results in that almost half of the elapsed simulation time is spent on waiting for heavily loaded processes handling $N_p/8$ Voronoi subdomains to accomplish their useful calculations. That involves Voronoi generators of less loaded subdomains into movement toward the adjacent heavy loaded ones, which leads to transfer of particles among them. After some balancing iterations, the decomposition reaches a~well-balanced geometry. At this moment both the waiting time for communication between the neighboring Voronoi subdomains and the elapsed time per a~simulation step reaches their minima (Fig.~\ref{fig:static}b). In such state, the numbers of particles in all subdomains are almost equal.

Figure~\ref{fig:static}c shows the evolution of the calculation speedup defined as a~ratio of the wall clock elapsed time $t^\infty_{32}$ per a~step in a~well-balanced final decomposition with 32 processes to the elapsed times measured during simulations using from 32 to 1024 processes. It is clear that the converging to a~well-balanced state takes more iterations for a~larger number of $N_p$. The~reason is that two types of Voronoi subdomains are initially generated in the test: those located near the boundary between the quarters where they are subjected to significant load imbalance with subdomains from other quarters, and the subdomains located inside each quarter, which are in a~good balance with surroundings. For a~larger number of $N_p$, a~ratio of the boundary subdomains to the inner ones becomes smaller, and more iterations are required to involve the internal subdomains in the balancing process.

It is worth to note that the Voronoi subdomain shapes transform to hexagonal during load balancing. It is because the hexagon has the lowest number of neighbors and a~ratio of its perimeter to area is the smallest among regular polygons, which allows to minimize the number of alien particles in boundary layers and thereby to reduce data transfer and communications between subdomains.

Final speedup that is shown in Fig.~\ref{fig:static}d is measured for well-balanced decompositions with using 32, 64, 128, 256, 512, and 1024 CPU cores. One can notice, that the speedup is almost linear up to 1024 cores. Final deviation from the perfect linear speedup for 1024 cores is caused by the increased ratio between the time required for data exchange and the useful time for particle simulation. The~reason for it is the growth of particle numbers in boundary layers between subdomains, which leads to a~growth of data exchange per a~simulation step. With the increase of boundary layer area and the corresponding decrease of inner area of subdomains, the data exchange masking by ``natives''--``natives'' calculations gets less efficient with $N_p$.

\section{Parallel performance in dynamic tests with materials in extremes}
\label{sec:dynamic_results}

In this section, the~performance of auto balancing algorithm is checked in the~modeling of complex material motion produces by extreme conditions, which leads to high-rate deformations and discontinuities in material, and thereby inhomogeneous spatial distribution of particles. In order to demonstrate the~obtained results clearly the two-dimensional rather than three-dimensional Voronoi decomposition is used as in the~previous section.

For the~first dynamic test of the~auto-balancing algorithm, the~ejection from a~rough metal surface produced by shock wave arrival is chosen \cite{Buttler2012,Dimonte2013,Shao2014,Durand2015}. Surface grooves on metal may have small sizes of the~order of $10$--$100$~$\mu$m depending on experimental setup. After reflection of a~flat shock from such surface, the~microscopic cumulative jets are formed, which may move with a~speed of several km/s.

The explosion of a~metal wire caused by fast energy deposition through the~ohmic heating is considered in the~second dynamic test. The~magnetic field produced by a~high current flowing in the~wire keeps it from the thermal expansion. The~external magnetic pressure is balanced by the~equal internal pressure of several GPa until the~current cutoff. Heating is accompanied by wire material evaporation increasing with temperature growth. At some moment the~electric breakdown of surrounding vapor happens, which results in switching of the~current into the~coronal plasma. After the~current cutoff in the~wire, the~magnetic pressure disappears, and the~highly-pressurized material of wire begins to expand freely. To perform an SPH simulation test for this expansion stage only, we use the~initial conditions provided by magnetohydrodynamic modeling of an aluminum wire~\cite{tkachenko2012study}.

Tremendous deformation and fragmentation of samples, observed in experimental conditions discussed above, challenges the~adaptability of any load-balancing algorithm with domain decomposition.

To close the~system of equations \eqref{SPH:Conti}--\eqref{SPH:Energy}, the~well-known Mie--Gruneisen equation of state is used to model materials in extreme states:
\begin{equation}\label{eq:mie_gruneisen_eos}
P - P_r(\rho) = \Gamma\rho[e-e_r(\rho)],
\end{equation}
where $\rho$ is the~density of material, $P$ is the~pressure, $P_r(\rho)$ is the~reference pressure, $e$ is the~specific internal energy per unit mass, $e_r(\rho)$ is the~reference specific internal energy, $\Gamma$ is the~Gruneisen parameter. We choose a~simple linear approximation of the~shock Hugoniot in the $u_s - u_p$ plane, where $u_s$ is a~shock wave velocity and $u_p$ is a~particle velocity behind the~shock front :
\begin{equation}
  u_s = c_a + s_a u_p,
\end{equation}
where $c_a$ and $s_a$ are the fitting parameters, which together with $\Gamma$ characterize the~specific material. Then this relation can be used to derive the~reference curves for pressure and internal energy as follows:
\begin{eqnarray}
\label{eq:mie_gruneisen_reference}
  P_r(x) &= &\rho_0 c_a{^2}\frac{1 - x}{[1 - s_a(1-x)]^2}\\
  e_r(x) &=& \frac{c_a{^2}}{2}\frac{(1 - x)^2}{[1 - s_a(1-x)]^2}
\end{eqnarray}
where $\rho_0$ is the~initial density and $x = \rho_0/\rho = 1 - u_p/u_s$ is a~compression ratio.

\subsection{Test 1: ejecta from grooved metal surface}

\begin{figure}[t!] 
  \center\includegraphics[width=0.9\linewidth]{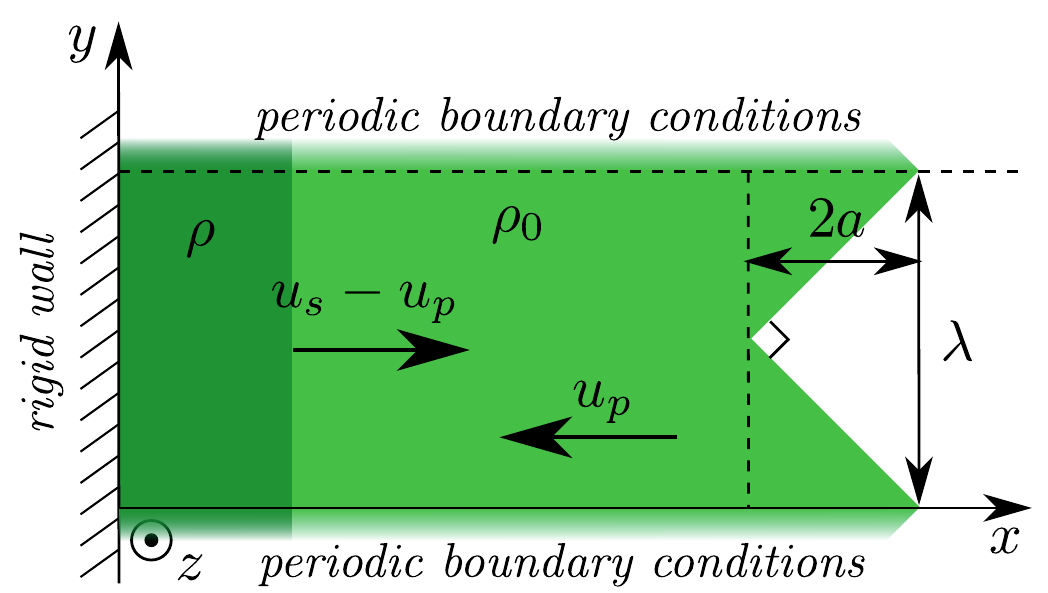}
  \caption{Geometry of lead sample used in simulation of ejecta from a~grooved surface. PBC are imposed along $y$- and $z$-axes. Impact on a~rigid wall with an initial particle velocity $-u_p$ generates a~shock wave propagating with velocity $u_s-u_p$ in the~chosen reference system toward the~grooved surface. }
  \label{fig:jets_setup}
\end{figure}

The ejection of a~cumulative jet from a~shock-loaded grooved metal surface is simulated for a~lead sample with a~90-degree opening angle of a~groove as illustrated by Fig.~\ref{fig:jets_setup}. The~periodic boundary conditions are imposed along the~axes $Oy$ and $Oz$. The~period (wavelength) of the~grooves $\lambda$ is set to $40$~$\mu$m, the~sample width is $2$~$\mu$m along $z$, and the~length is $140$~$\mu$m. The~size of SPH particles is set to $0.08$~$\mu$m, so the~number of particles in this modeling is 20.8~millions. Parameters for the~equation of state~\eqref{eq:mie_gruneisen_eos} are: $\rho_0 = 11.35$~g/cc, $c_a = 2.58$~km/s, $s_a = 1.26$, and $\Gamma = 1.7$.

Initially, all particles are set to velocity $u_p=1$~km/s along the~direction $x$ toward a~steady piston in the~form of a~flat rigid wall. It results in the~collision with the~piston, which stops the particles at the~wall and generates a~shock wave propagating with the~velocity of $u_s-u_p$ in the~chosen coordinate system in which the~piston/wall is at rest.

\begin{figure*}[t]   
  \center\includegraphics[width=0.7\linewidth]{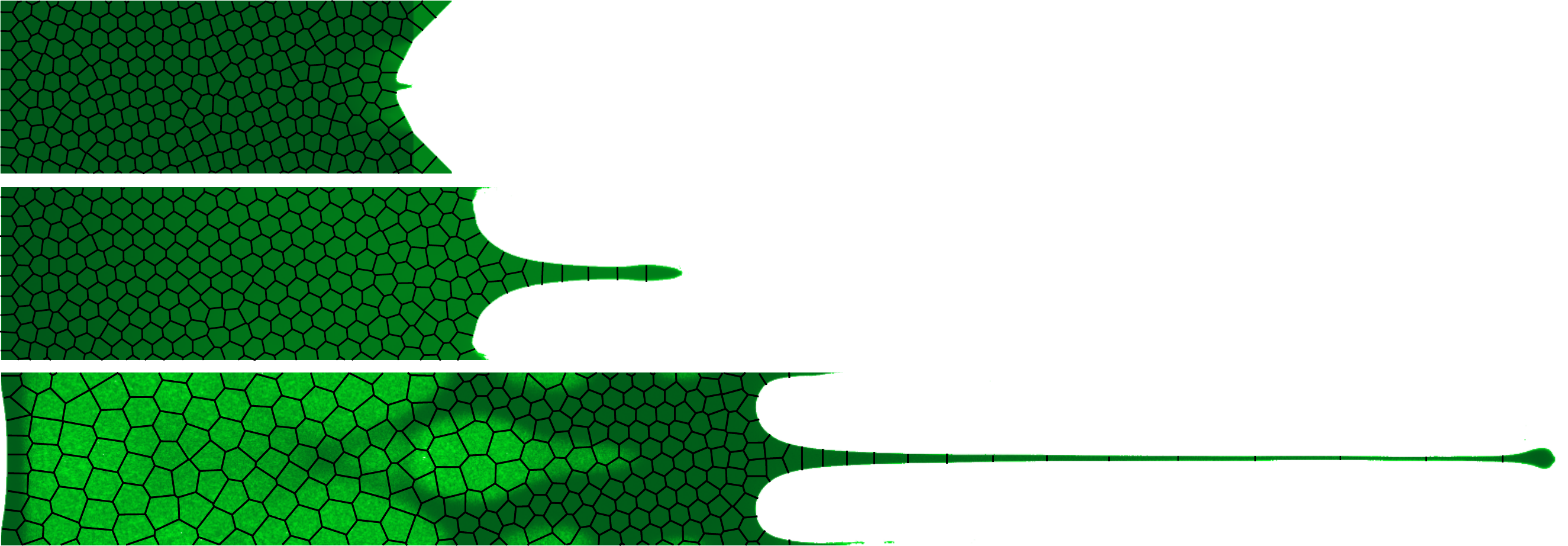}
  \includegraphics[width=0.29\linewidth]{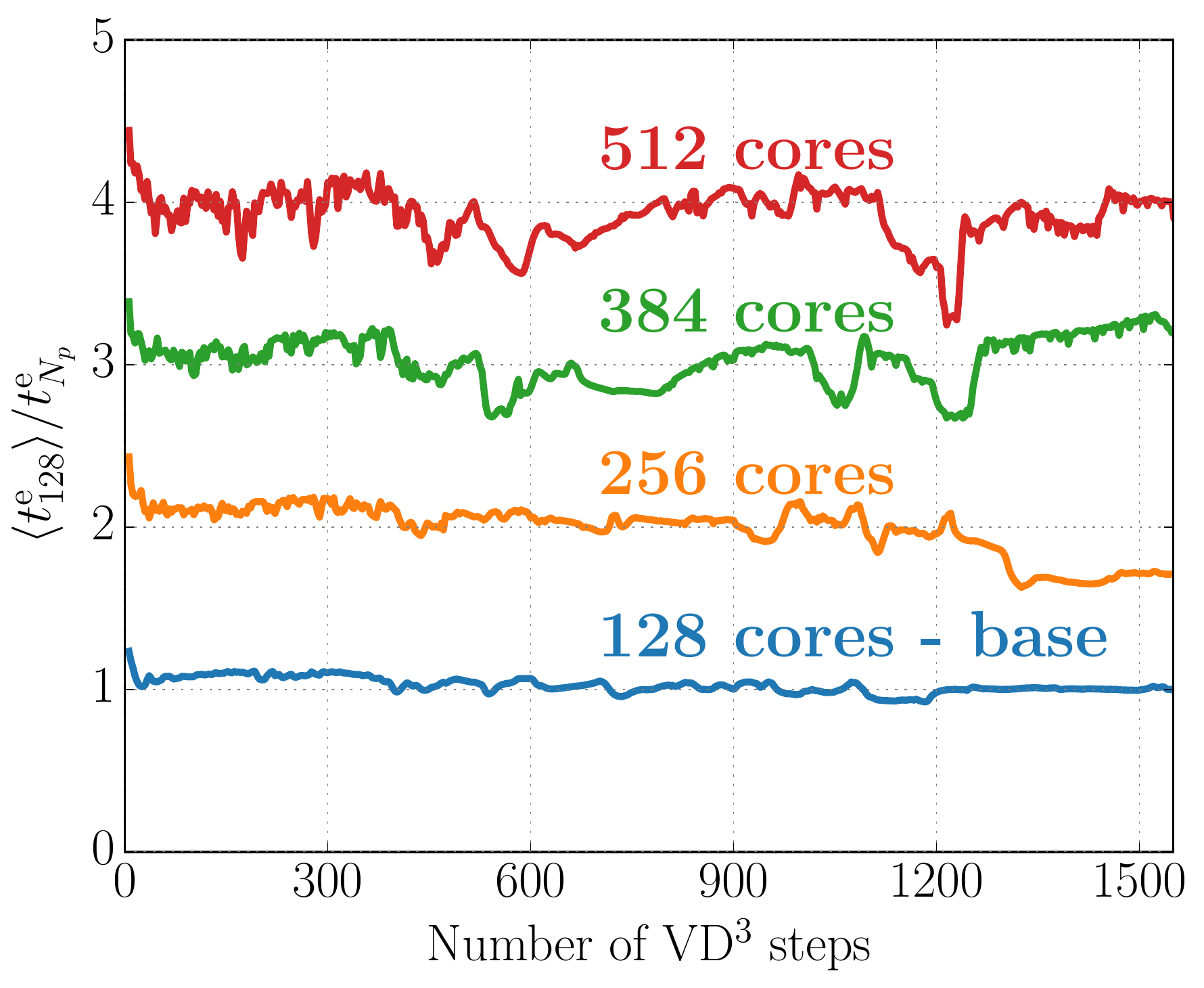}
 \caption{Maps of Voronoi subdomain boundaries and mass distributions in ejecta obtained in simulation using the \VD on 256 processors is shown on the~left, where the~darker blue color corresponds to higher particle density. The~relative speedups $\langle t^e_{128}\rangle/t^e_{N_p}$ as functions of simulation step in modeling the~same problem with 128, 256, 384, and 512 cores are presented on the~right. Averaged elapsed time per step in modeling with 128 cores $\langle t^e_{128}\rangle$ is used for evaluation of the~relative speedup. }
  \label{fig:jets_speedup}
\end{figure*}

Figure~\ref{fig:jets_speedup} shows the~evolution of mass distribution with the~corresponding Voronoi diagram in simulation on 256 processor cores. During the~shock compression, the~areas of Voronoi subdomains becomes smaller. The~jet formation is accompanied by an inflow of Voronoi subdomains into the~jet. During rarefaction, there appear dense and low-density regions in the~material sample, which are appropriately covered by the~Voronoi diagram. Darker regions with larger numbers of particles are covered by smaller Voronoi subdomains, while the~regions with lesser dense material are covered by larger subdomains. The~simulation speedup with the~increase of processor number remains almost linear despite large variations of mass distribution during the~test. 

To test the~strong scalability, simulations are performed using 128, 256, 384 and 512 processor cores. The~plot of the~obtained speedup is shown in Fig.~\ref{fig:jets_speedup} on the~right. The~wall clock elapsed time required for one simulation step decreases with the~increase of the~number of processes.

\begin{figure*}[t]   
  \includegraphics[width=0.7\linewidth]{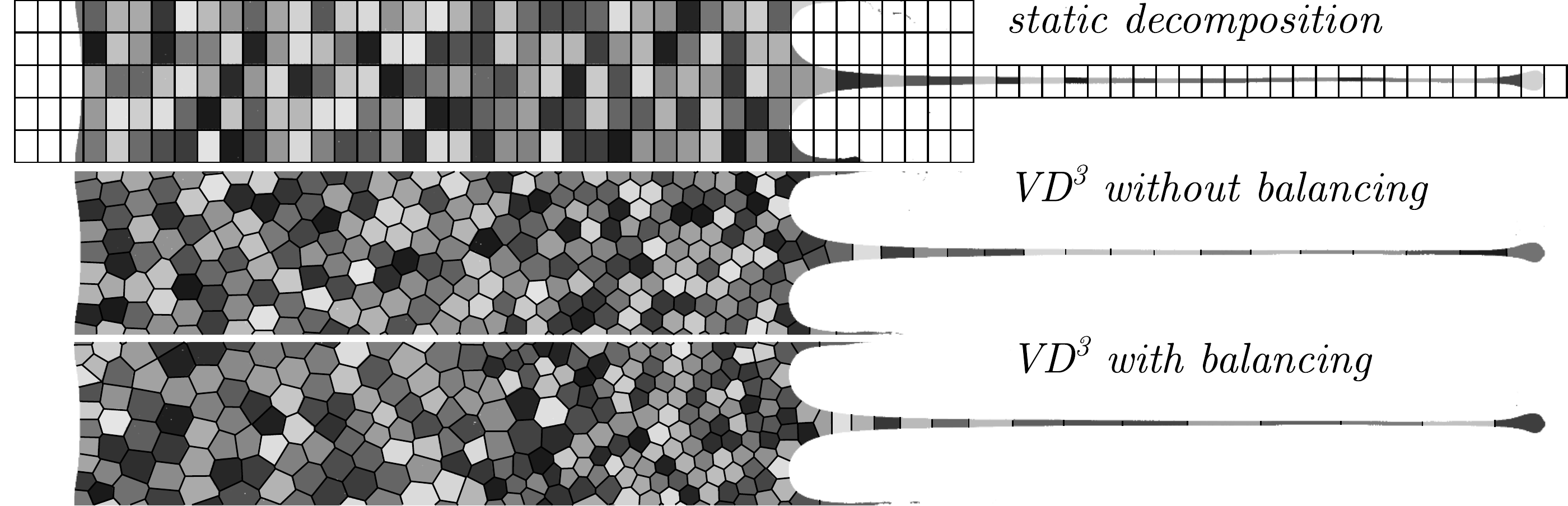}
  \includegraphics[width=0.29\linewidth]{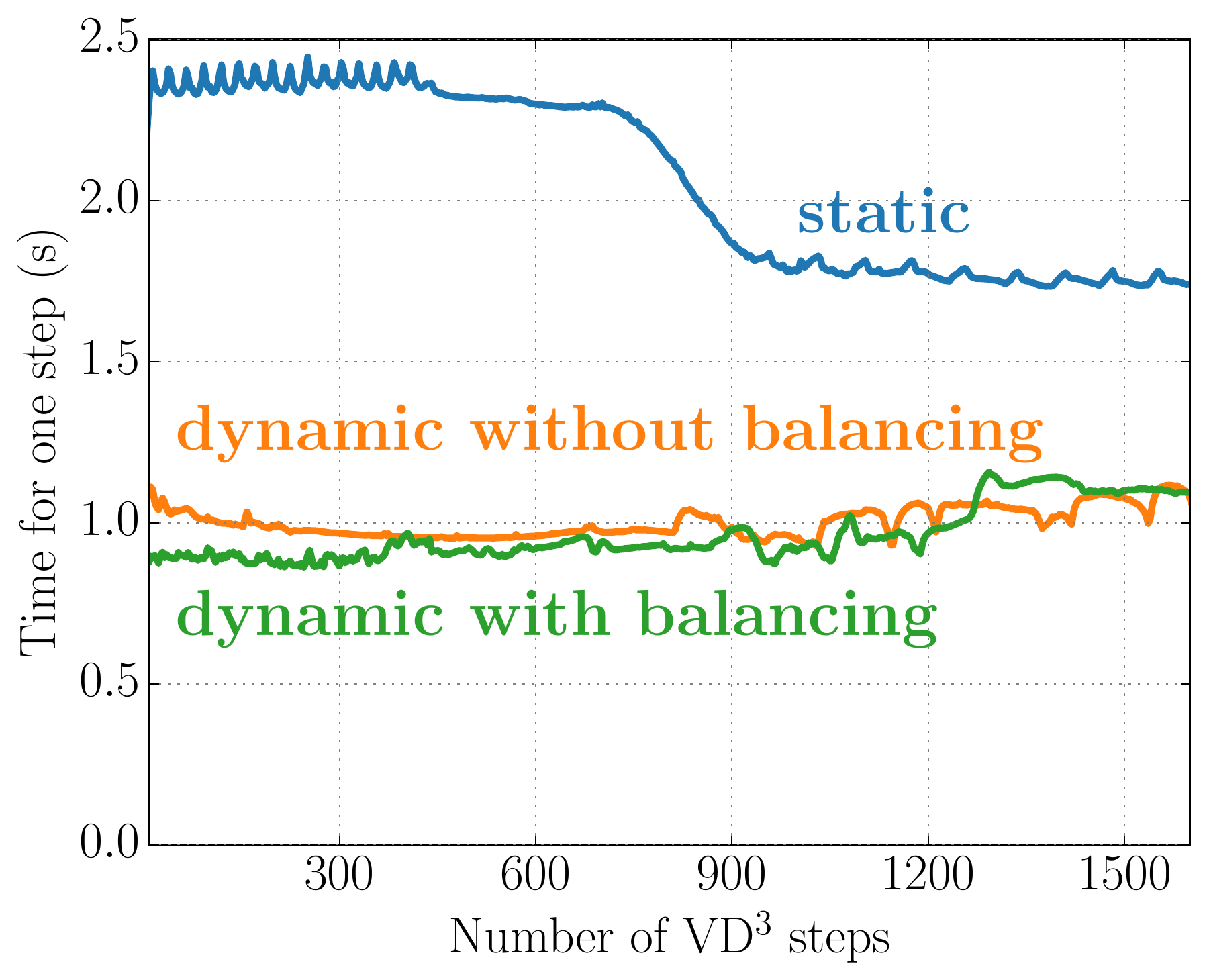}
 \caption{Maps of subdomain boundaries obtained in simulations of ejecta on 256 cores using the~static rectangular decomposition, dynamic Lagrangian without balancing and \VD with load balancing, from left to right respectively. Shades of gray are used to distinguish different subdomains only. The~right plot shows the~corresponding elapsed times per a~step during simulation.
}
  \label{fig:jets_decomp_compare}
\end{figure*}

To demonstrate the~advantages of dynamic Voronoi decomposition, a~comparison with a~static domain decomposition is performed as shown in Fig.~\ref{fig:jets_decomp_compare}. To do this, 255 rectangular subdomains are placed to cover a~large computational domain, which size is determined by possible material motion during simulation of ejecta. The~initial sample occupies less than half of these steady subdomains, which leads to an increase in the~elapsed time required for one simulation step due to uneven distribution of computational loads between processes, see Fig.~\ref{fig:jets_decomp_compare}. After formation of a~jet and expansion of the~sample toward the~unoccupied subdomains, the~elapsed time per a~step decreases.

The neighborhood reassignment of Voronoi subdomains is always performed after each balancing iteration in a~dynamical decomposition mode of our code. Since the balancing displacement is limited by the minimal subdomain size the neighbor subdomain lists do not change much after each iteration and the reassignment takes a~short time required for communications only between a~central Voronoi subdomain and subdomains in the first and second coordinate sphere surrounding the central one. In a~static decomposition mode the neighbor subdomain lists are prepared only once at the initial decomposition procedure. Also our code skips the balancing calculations in the static decomposition mode. Thus the code in this mode avoids the procedures required for the dynamical decomposition mode.

Turned off balancing means that $\theta = 1$ in Eq.~\eqref{eq:TotalOffset}, at which
the~Voronoi diagram generators coincide with the corresponding geometric centers Eq.~\eqref{eq:VoronoiGeomCenters}. By contrast, both dynamic decompositions with and without load balancing maintain this time per a~step at nearly the~same level during the~entire simulation. But the~usage of load balancing results in lesser total simulation time.

It is worth noting that the~described algorithm realizes the~local load balancing strategy only. In modeling of a~system where a~large elongation happens, such as in jetting, the~connectivity of the~individual parts of the~diagram decreases to one or two links per a~Voronoi subdomain. Then it takes a~long time to transfer information about long-range imbalances in the~diagram to push additional worker processes to the~highly loaded subdomains. As a~result, many balancing iterations can be required to establish a~global well-balanced decomposition.

To reduce a~global load imbalance faster, some global balancing algorithm must be utilized. In particular, the~use of a~guided drift of Voronoi generators in the~direction toward the most  overloaded subdomain makes it possible to improve the~global load balance for a~shorter time. This approach can also help if a~large number of particles is created in some subdomain during the~simulation. Such global balancing algorithm using the drift displacement in addition to the locally determined displacements demonstrate a~notable acceleration of balance convergence in our preliminary tests, but it still requires the fine adjustment to minimize the undesired interference with the local balancing algorithm.

\subsection{Test 2: explosion of aluminum wire}\label{subsec:wire}

To test how the auto balancing algorithm copes with the fast two-dimensional expansion of material the explosion of aluminum wire is simulated by our SPH code. At the~beginning a~cylinder of radius $R= 6$~$\mu$m and height $h_z=R/3$ is assembled using 3.6 millions of aluminum SPH-particles with the size of $d_i=40$~nm. The~cylinder surface of wire remains free in the $x$--$ y$ plane, and periodic boundary conditions are imposed along the $Oz$ axis.

The reference curves which are used in the~shock Hugoniot for aluminum~\eqref{eq:mie_gruneisen_reference}, are given in accordance with the molecular dynamics data~\cite{zhakhovsky2012cavitation}: $\rho_0 = 1.593$~g/cc, $c_a=3$~km/s, $s_a=1.85$, $\Gamma=1.5$. Magnetohydrodynamic modeling \cite{tkachenko2012study} of the aluminum wire demonstrated that because of the ohmic heating the material melts and expands much, but the external magnetic pressure of $P=2.8$~GPa stops this expansion until the electric current cutoff happens. The~cutoff time is used as an initial zero time in the SPH simulation. The~density of $\rho_0 = 1.593$~g/cc and temperature of 4000 K established at this moment are used as the initial conditions in this SPH simulation test. The~tensile strength of $\sigma_c$ in the hot molten aluminum turned out to be equal to $1.9$~GPa, which corresponds to the density drop to $\rho=0.9\rho_0$ according to the~results of MD simulation~\cite{zhakhovsky2012cavitation}.

\begin{figure*}[t]     
  \includegraphics[width=0.7\linewidth]{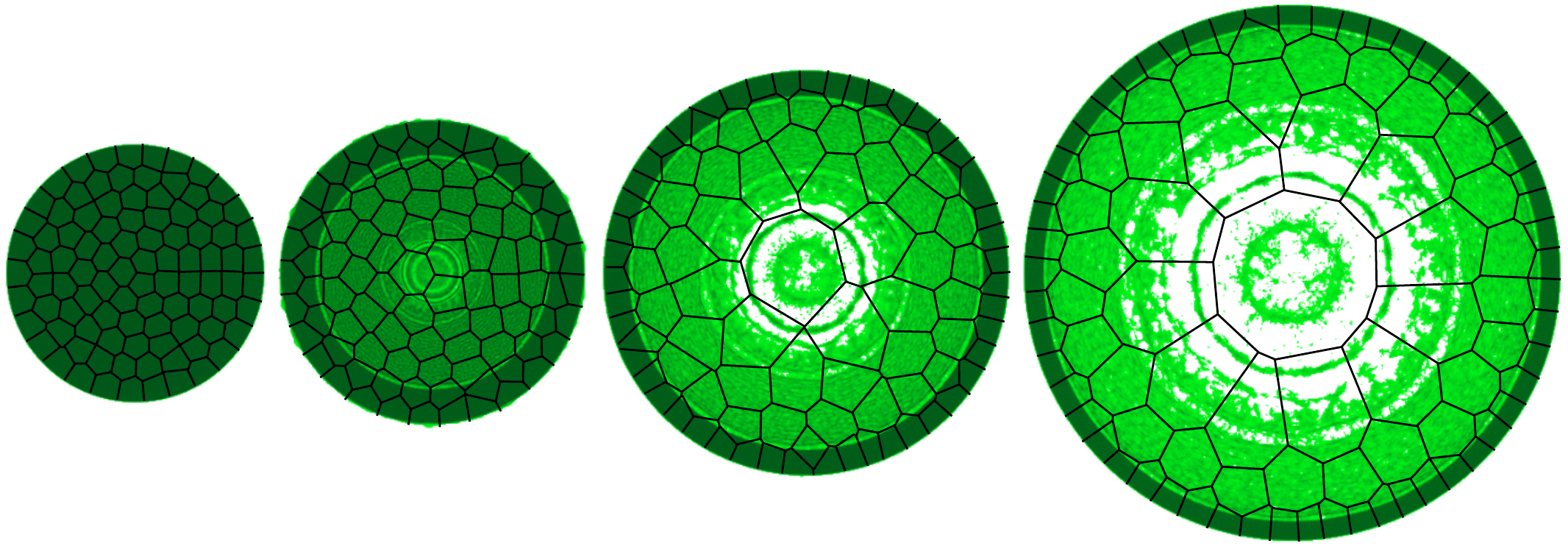}
  \includegraphics[width=0.29\linewidth]{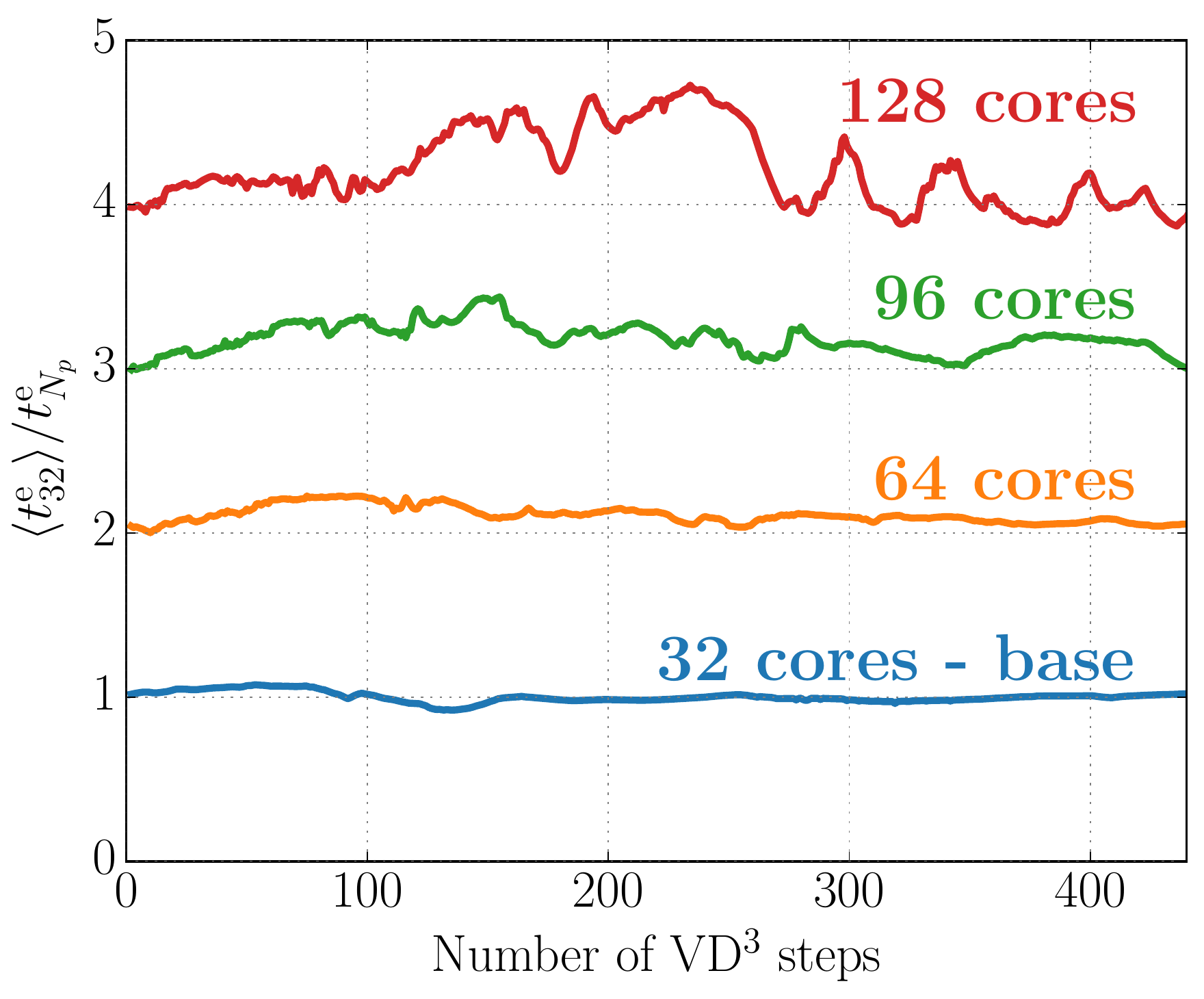}
  \caption{Maps of Voronoi subdomain boundaries and mass distributions during wire explosion simulated with using the \VD on 96 processor cores is shown on the~left, where the~darker blue color corresponds to higher particle density. The~relative speedups $\langle t^e_{32}\rangle/t^e_{N_p}$ as functions of simulation step in modeling the~same problem with 32, 64, 96, and 128 cores are presented on the~right. Averaged elapsed time per step in modeling with 32 cores $\langle t^e_{32}\rangle$ is used for evaluation of the~relative speedup. }
  \label{fig:wire_density_evolution}
\end{figure*}

Figure~\ref{fig:wire_density_evolution}, on the~left, shows evolution of mass distribution and Voronoi decomposition in the cross-section of wire during its expansion in SPH simulation using \VD on 96 processor cores (1 master process for input-output and 95 workers). With time $t>0$, a~rarefaction wave moves radially from the cylinder surface to the wire axis. During the~convergence of the~rarefaction wave to the~axis, the~material density drops below the break condition $\rho < 0.9\rho_0$ at the depth of $1$~$\mu$m from the wire surface, at which the interaction between SPH-particles is ceased, and many small voids appear within a~central part of the wire. As a~result of simultaneous void formation at the same depth of $1$~$\mu$m, a~dense liquid shell with this thickness is detached from the wire, while the internal part of wire undergoes farther expansion and fragmentation.

The elapsed times per a~simulation step are tracked in SPH simulations of wire explosion using 32, 64, 96 and 128 processor cores to calculate the performance speedup. The~reference elapsed time is assume to be an averaged time per a~step $\langle t^e_{32}\rangle$ obtained for entire simulation using 32 cores. Then the relative speedup $\langle t^e_{32}\rangle/t^e_{N_p}$ as function of simulation step is plotted for all used $N_p$ on the right of Fig.~\ref{fig:wire_density_evolution}.

Initially, all Voronoi subdomains cover equal parts of the~wire in terms of the~number of particles. During expansion, a~region of low density is formed in the~central part of the~exploding wire, and the~balancing algorithm drives most of the~Voronoi subdomains to the~dense shell. The~inner part of the~wire is covered by the large Voronoi subdomains in accordance with the~low particle density there. However there is an opposite geometrical constrain which requires to cover whole expanding cylinder by Voronoi subdomains involved in simulation. As a~result the number of subdomains covering the thin dense shell begins to decrease when more subdomains need to be used to cover the huge area of the expanding inner part. Then the loads of shell subdomains start to increase slowly, while the load of inner subdomains decreases. The~balancing algorithm with the fixed number of subdomains cannot improve the imbalance produced by such geometrical constrain. As a~result, the elapsed time grows slowly at later times of wire explosion.

\begin{figure*}[h]      
  \includegraphics[width=0.7\linewidth]{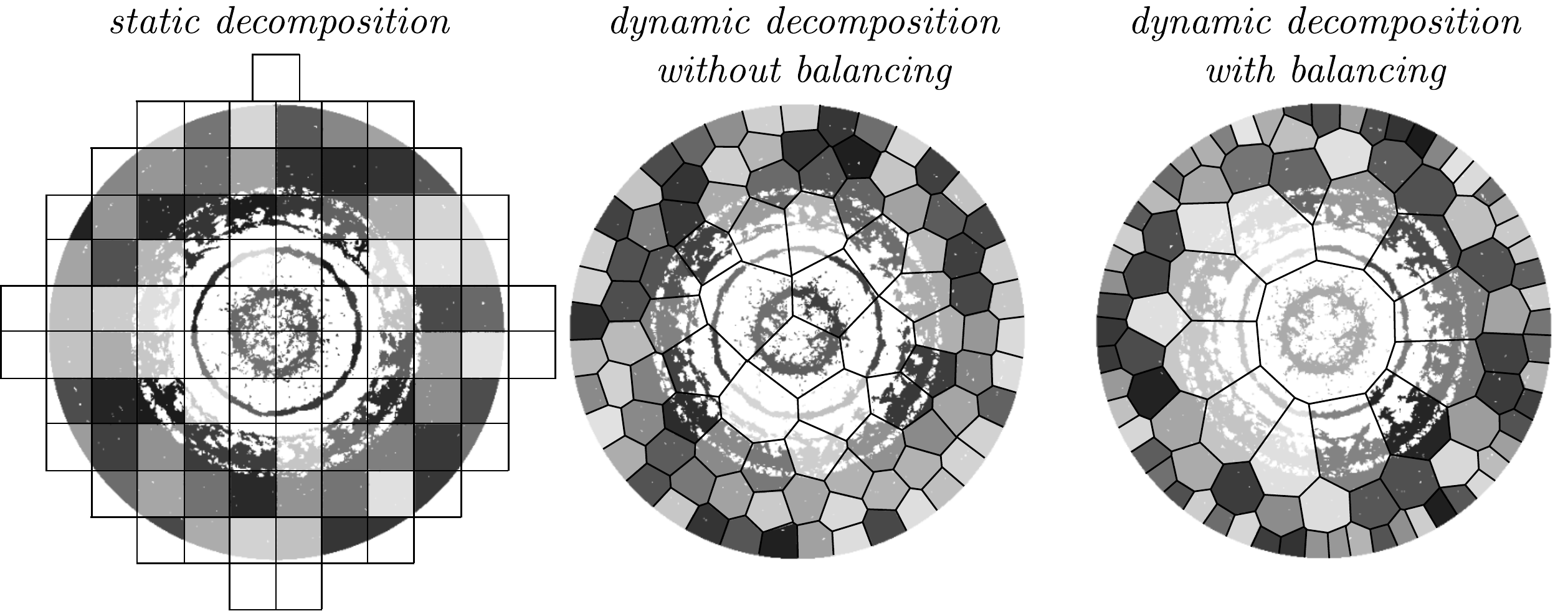}~
  \includegraphics[width=0.29\linewidth]{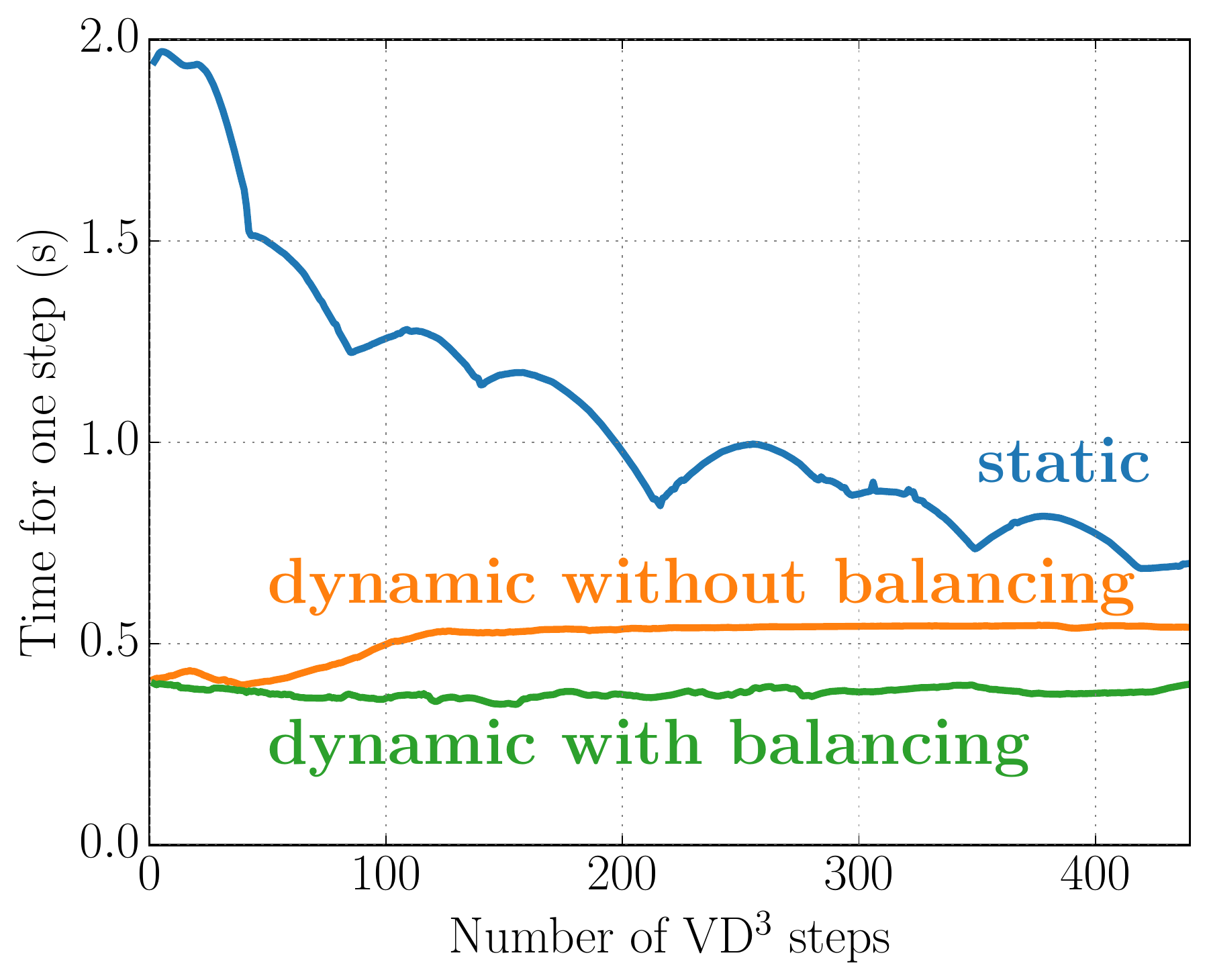}
  \caption{Maps of subdomain boundaries obtained in simulations of wire explosion on 96 cores using the~static rectangular decomposition, dynamic Lagrangian without balancing and \VD with load balancing, from left to right respectively. Shades of gray are used to distinguish different subdomains only. The~right plot shows the~corresponding elapsed times per a~step during simulation. The~areas of Voronoi subdomains become smaller on a~liquid cylindrical shell with the~highest density to reduce the~computational loads in \VD with load balancing. }
  \label{fig:wire_decomp_compare}
\end{figure*}

Comparison of static rectangular and dynamical Voronoi decompositions without load balancing and with balancing with 96 processor cores are presented in Fig.~\ref{fig:wire_decomp_compare}. Static decomposition is prepared at the initial decomposition with Voronoi generators located on a~regular mesh. The~rectangular size is chosen to cover a~large area which has enough room for the expanded wire material at the end of simulation. The~time for calculating one simulation step for all decomposition types is shown on the left frame in Fig.~\ref{fig:wire_decomp_compare}.

Only 24 cores/subdomains from 95 are filled by particles at the beginning of simulation using the static decomposition. As a~result, here the elapsed time is five times longer than it is obtained in simulations using the dynamical Voronoi decompositions, where all 95 subdomains are filled by particles. With expansion of wire the SPH particles begin to enter into other rectangles of the static decomposition, which results in a~more even distribution of the load, so the elapsed time decreases. However, this time is still greater than the times from the both Voronoi decompositions. Decomposition without load balancing has less mobile Voronoi subdomains striving to preserve their areas, which deteriorates the load balance with expansion. Thus, the dynamic Voronoi decomposition with load balancing is superior to other decompositions in this test. Moreover, it provides a~stable elapsed time per a~step throughout the entire simulation due to a~more even use of all processors.

For the~dynamic decomposition without the usage of balancing displacements ($\theta = 1 $ in Eq.~\eqref{eq:TotalOffset}), the~Voronoi generators are strictly set to subdomain geometric centers~\eqref{eq:VoronoiGeomCenters}. Such decomposition results in less mobile Voronoi subdomains, which increases the~elapsed time of simulation as seen in Fig.~\ref{fig:wire_decomp_compare}.
Both dynamic Voronoi decompositions give a~stable time step throughout the~simulation due to the~more even use of the~resources of all processors, but the usage of load balancing results in the~better positions of Voronoi subdomains, providing the~shortest time for a~single step. 

\begin{figure}[t]   
  \center\includegraphics[width=0.9\linewidth]{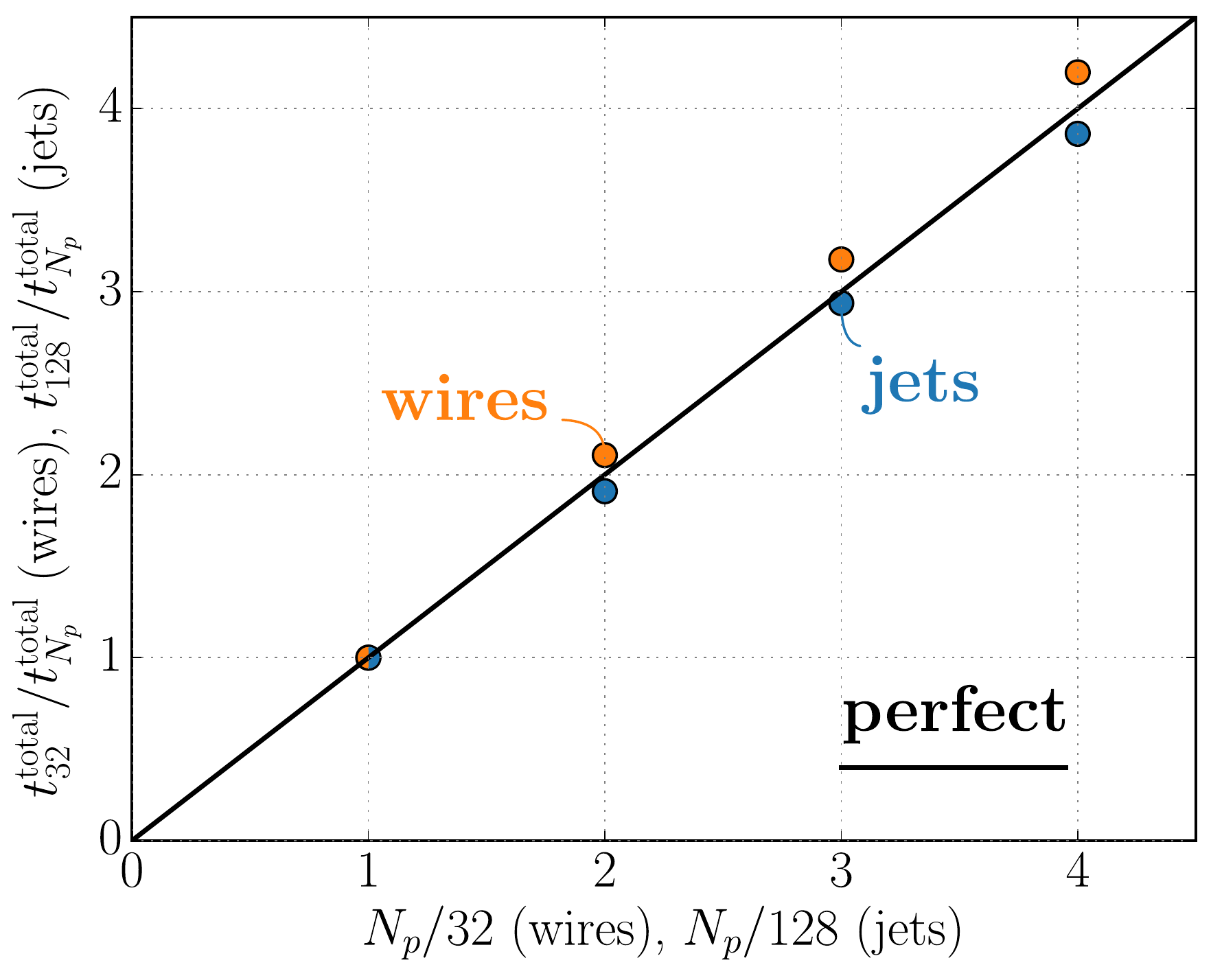}
 \caption{Strong scalability for 1st and 2nd dynamic tests of the~load balancing algorithm. Here the~wall clock elapsed times $t^e_{128}$ and $t^e_{32}$ of entire modeling are chosen as reference values, for the~1st test with ejecta and the~2nd test with wire explosion respectively. The~strong scalability is close to the~perfect linear speedup in both test, but the~2nd test demonstrates a~slightly superlinear speedup because the~major part of wire mass happens to be allocated in a~thin shell during the~explosion, which corresponds to transition from 2D to 1D decomposition for larger numbers of processes $N_p$ involved. }
  \label{fig:wires_jets_final_strong_scalability}
\end{figure}

This 2nd test of the \VD algorithm applied for an exploding wire shows not only better parallel efficiency in comparison with the static decomposition method, but also a~strong linear scalability shown on Fig.~\ref{fig:wires_jets_final_strong_scalability}. This Figure also demonstrates the superlinear speedup which can be achieved in this test in contrast to the speedup obtained in the 1st ejecta test. We think that the additional acceleration is caused by transition from a~two-dimensional initial decomposition to a~quasi-one-dimensional decomposition of the thin dense shell which is seen in Fig.~\ref{fig:wire_density_evolution}. The~Voronoi subdomains on the shell have the most occupied boundary layers only with two neighbor subdomains, which minimizes the particle data exchange and makes more efficient data exchange masking. 

\section{Conclusion}

When using particle methods for modeling phenomena specific for high energy density physics, the variety of particles representing the material samples can evolve from an initially homogeneous distribution (for example, a~set of regularly packed particles equal in mass and volume) into a~highly heterogeneous one varied in the spatial arrangement and the particle characteristics. The~major reasons for such evolution are:
\begin{itemize}
\item particle compression in shock waves and expansion in rarefaction waves;
\item nonuniform acceleration of material flow by nonplanar compression and rarefaction waves;
\item collision between material flows and formation of cumulative jets;
\item formation of cavities and free boundaries of complex shape if material is fragmented;
\item adaptive merging or splitting of particles to improve the accuracy of modeling.
\end{itemize}

All listed reasons cause the significant load imbalance and reduce the computational efficiency of parallel modeling of material motion. To address this issue, we have developed the efficient automatic load balancing algorithm for parallel modeling on the large computing clusters with distributed memory. It is based on the spatial dynamic decomposition of simulated samples between Voronoi subdomains, where each subdomain is handled by a~single computational process. Voronoi subdomains are allowed to gradually change their shape and position in order to reduce the local imbalance of computational loads via the re-balancing transfer and redistribution of particles between the neighbor processes.

Adaptive load balancing and high computational efficiency are the main advantages of this algorithm. Since the generators of Voronoi subdomains can move freely, the resulting well-balanced Voronoi diagram provides the most computationally attractive coverage for the simulated samples. The~evolution of the diagram during simulation is guided by the natural Lagrangian motion of material combined with the load balancing displacements. The~algorithm demonstrates fast convergence in static tests of a~stationary sample with load imbalance imposed initially and almost linear strong scalability for processor core numbers from tens to several thousand.

Parallel efficiency of our code is illustrated by modeling materials in extreme conditions characterized by large pressure and velocity gradients, at which the spatial distribution of particles can vary greatly in time. In such conditions, our algorithm performs as efficiently as in the static tests and provides the similar strong scalability. The~achieved parallel efficiency is superior to that provided by static decomposition algorithm in all dynamic tests.

Our algorithm is recommended to be used in simulations of material samples with essentially dynamic boundaries, flows with large relative velocities, and particles described by physical models with notably different computational costs, where the automatic balancing with the stable high load of all processes is required during simulation.

\section{Acknowledgement}
This work is supported by the Russian Science Foundation grant 14-19-01599.

\end{document}